\documentclass[journal,compsoc]{IEEEtran}
\ifCLASSINFOpdf
\else
\fi
\usepackage{amsfonts,amsmath,amssymb}
\usepackage{mathtools}
\usepackage{mdwmath}
\usepackage{cuted}
\usepackage{mdwtab}
\usepackage{amsthm}           
\usepackage{bm}               
\usepackage{units}            
\usepackage{graphicx}
\usepackage{epstopdf}
\usepackage[linesnumbered,ruled,vlined]{algorithm2e}

\usepackage{url}

\usepackage{paralist}
\usepackage{color}
\usepackage{authblk}

\newcommand{\Fig}[1]{Fig.~\ref{fig:#1}}

\newcommand{\Eq}[1]{Eq.~(\ref{eq:#1})}

\newcommand{\Lc}{\mathcal{L}}

\newcommand{\Rb}{\mathbb{R}}
\newcommand{\Rs}{\mathbb{R}^2}

\newcommand{\Ed}{\mathbb{E}}

\newcommand{\Pd}{\mathbb{P}}
\newcommand{\sinr}{\mathrm{SINR}}

\newcommand{\dr}{\mathrm{d}}
\newcommand{\Rcone}{u(\omega,\gamma)}
\newcommand{\Ru}{v_i(\gamma_i,\phi_{ij})}
\newcommand{\Rl}{w_i(\gamma_i,\phi_{ij})}

\newcommand{\Upper}{u}
\newcommand{\Lower}{l}


\DeclarePairedDelimiter\floor{\lfloor}{\rfloor}

\begin{document}
%
\title{A Stochastic Geometry Model of Backhaul and User Coverage in Urban UAV Networks}
%
%
%

\author{Boris Galkin,
        Jacek~Kibi\l{}da,
        and~Luiz~A. DaSilva
}

\affil{CONNECT, Trinity College Dublin, Ireland \\
\textit{E-mail: \{galkinb,kibildj,dasilval\}@tcd.ie}}

\maketitle

\begin{abstract}
Wireless access points on unmanned aerial vehicles (UAVs) are being considered for mobile service provisioning in commercial networks. To be able to efficiently use these devices in cellular networks it is necessary to first have a qualitative and quantitative understanding of how their design parameters reflect on the service quality experienced by the end user. In this paper we model a network of UAVs operating at a certain height above ground to provide wireless service within coverage areas shaped by their directional antennas, with the UAVs using the existing terrestrial base station network for wireless backhaul. We provide an analytical expression for the coverage probability experienced by a typical user as a function of the UAV parameters. Using our derivations we demonstrate the existence of an optimum UAV height which maximises the end user coverage probability. We then explore a scenario where the UAVs adjust their individual heights to meet their backhaul requirements while at the same time attempting to maximise the coverage probability of the end user on the ground.
\end{abstract}


%
\IEEEpeerreviewmaketitle

\section{Introduction}
To meet growing data demands and support emerging services, telecommunications operators and over-the-top service providers are considering the use of unmanned aerial vehicles (UAVs) for delivering wireless connectivity. These wireless-provisioning UAV platforms vary greatly in size and operating range, with high-altitude UAVs operating across hundreds of kilometers at altitudes previously reserved for manned aircraft on one end, and miniature quadcopter-style UAVs with ranges of a few hundred meters on the other. UAVs in the latter category are in particular drawing the attention of the internet of things (IoT) community: the authors of \cite{Motlagh_2016} suggest that the majority of UAVs in IoT applications will be miniature devices that operate at heights below 300 meters. The reason for this is that miniature, low altitude UAVs offer lower cost, more flexible deployment and they make use of airspace which is far less utilised by manned aircraft and is therefore subject to more relaxed regulations \cite{Atkins_2010}. 

Whereas a terrestrial base station (BS) may use a wired backhaul into the core network, UAVs require a wireless channel for their backhaul. This wireless backhaul may take the form of a free space optical (FSO) or millimeter wave (mmWave) connection to a dedicated ground station \cite{Alzenad_2016}\cite{Xiao_2016}; alternatively, it may involve the UAVs opportunistically connecting to infrastructure originally designed for serving ground user equipment. The 3GPP has begun a study on the feasibility of using LTE BSs to deliver wireless connectivity to low-altitude UAVs. The authors of \cite{Lin_2017} analyse BS-UAV channel performance in a rural environment and conclude that UAVs are able to receive adequate service from terrestrial BSs in the presence of interference sources, even when BS antenna downtilt causes attenuation to the signal between the BS and the UAV.

 For the UAV network to deliver reliable service to the end user on the ground, the UAVs must optimise their connectivity to the end user while simultaneously meeting their own wireless backhaul requirements. A low-altitude UAV network may be operating above a built-up urban area or below building heights in so-called urban canyons. These environments will play a significant role in the performance of the UAV network.
 
 This paper extends our work in \cite{Galkin_2017} to include UAV backhauling into the core network through a network of LTE BSs. We employ stochastic geometry to model the effect of the urban environment on the ability of a UAV network to establish a wireless backhaul through an existing terrestrial network and the resulting coverage probability that is experienced by the end user of the UAV network. Our model takes into account parameters such as building density, UAV antenna beamwidth and BS antenna downtilt, and can represent different wireless fast-fading behaviours through generalised Nakagami-$m$ fading. As we consider actual building height distributions, our model allows us to analyse UAV deployments with cell sizes comparable to terrestrial picocells, with UAVs hovering at heights of around 100 meters and exhibiting coverage ranges on the order of a couple of hundred meters. We demonstrate that, for a given beamwidth of the UAV antenna, there exists an optimum height which maximises the coverage probability for a given UAV density and SINR threshold. We also demonstrate how good end user coverage can be achieved when the individual UAVs in the network are permitted to select their own height based on their individual backhaul as well as the system-wide user coverage probability. To our knowledge, we are the first to propose a stochastic model for a UAV network which considers both the UAV-user link and the UAV backhaul link simultaneously, with both links being affected by interference and LOS-blocking effects from the environment. 
 
 This paper is structured as follows. In Section II we describe the related work that models UAV network performance. In Section III we outline our system model and describe the end user and backhaul wireless channels. In Section IV we provide a mathematical analysis of the coverage probability experienced by the end user. In Section V we generate numerical results via simulations and discuss the various trade-offs and performance impacts that arise in our model. We conclude in Section VI.

\section{Related Work}
The wireless community has published several works on the modelling of wireless links between the UAVs and terrestrial devices. In \cite{Matolak_2015} the authors report the results of a measurement campaign to characterise the pathloss and multipath effects of the air-to-ground channel in a variety of environments for the L and C wireless bands, for altitudes in the range of \unit[500-2000]{m}. In \cite{Simunek_2013} the authors consider a low-altitude UAV operating in an urban environment: using field measurements and statistical analysis, they demonstrate that the air-to-ground channel in the \unit[2]{GHz} band behaves similarly to that of land mobile satellite signals. The authors of \cite{Goddemeier_2015} investigate the properties of the air-to-air channel between UAVs using an IEEE 802.11n link. From field measurements, the authors describe the multipath fading effect of the channel at different altitudes using a Rice model. In \cite{Cai_2017,Amorim_2017} the authors consider low-altitude UAVs operating in a suburban environment; they characterise the wireless channel from an LTE BS to a UAV at different altitudes and demonstrate the gain in received power that occurs as the UAVs increase their altitude. The authors of \cite{Al-Hourani_2017} consider a similar scenario and measure the channel pathloss in terms of the vertical angle between a UAV and its corresponding BS. They demonstrate that there exists a tradeoff in the channel performance as the vertical angle increases, due to the simultaneous impact of the improved LOS conditions and the deteriorating antenna gain due to BS antenna downtilt. In \cite{AlHourani_20142} the authors use an ITU model to describe an urban environment and then apply raytracing simulations to describe the pathloss from LOS and non-line-of-sight (NLOS) components. The same authors in \cite{AlHourani_2014} model the probability of a UAV having an LOS channel to a user as a sigmoid function of the vertical angle between the UAV and user; they then use this model to describe the coverage radius of the UAV as a function of pathloss and demonstrate how the UAV height can be optimised to maximise the coverage radius in an interference-free environment. 

The sigmoid LOS model proposed in \cite{AlHourani_2014} has subsequently seen widespread use for simulating UAV networks. In \cite{Mozaffari_2015, Hayajneh_2016} the authors use this sigmoid function LOS model to optimise UAV height for different performance metrics, and in \cite{Mozaffari_20162, Fotouhi_2016} the authors apply multi-objective optimisation to UAV networks, using the sigmoid LOS model to characterise the received signal strength. Note that in the above works the UAV locations are assumed to either be known $a$ $priori$ or are found as part of an optimisation problem.

 Stochastic geometry is an alternative method for modelling the spatial relationships in a UAV network. Without prior knowledge of the UAV locations, it is possible to describe the UAVs as being distributed in space randomly, according to a point process. This approach is followed in \cite{Ravi_2016} and \cite{Chetlur_2017}, in which the authors derive the coverage probability for a stochastic UAV network under guaranteed LOS conditions for a fading-free and a Nakagami-$m$ fading channel. The authors describe a fixed number of UAVs operating within a fixed area at a certain height above ground and demonstrate how an increase in height results in a decrease in the coverage probability. In \cite{Chetlur_2017} the authors demonstrate how larger values of fading parameter $m$ reduce the variance of the random signal-to-interference ratio (SIR) experienced by the user. Stochastic geometry is applied in \cite{Zhang_2017} to optimise UAV density in a radio spectrum sharing scenario under guaranteed LOS conditions. In our own previous work \cite{Galkin_2017} we considered a UAV network with directional antennas, with channels affected by Nakagami-$m$ fading and a distance-dependent LOS model. We derived an expression for the coverage probability as a function of UAV height, density and antenna beamwidth and demonstrated how there exists an optimum height for a given UAV density and beamwidth that maximises the user coverage probability. Our conclusions are corroborated by the subsequent work in \cite{MahdiAzari_2017}, where the authors analysed a UAV network under lognormal fading conditions.
 
 The UAV backhaul link has received less attention from the wireless community than the user link. The works cited above assume that a wireless backhaul is available to the UAVs at all times and limit the scope of their work to the user link. Examples of papers which explicitly include the UAV backhaul constraint as part of their system model include \cite{Kalantari_2017}, \cite{Kalantari_20172} \cite{Chen_2017}, \cite{Zeng_2016} and \cite{ZhangZeng_2017}. In \cite{Kalantari_2017} the authors optimise the location of a UAV to maximise throughput for a group of users in an interference-free environment, subject to a backhaul constraint. They extend this work in \cite{Kalantari_20172} to consider interference in the UE-UAV link. The authors assume the backhaul uses a dedicated band, so no interference in the backhaul link. In \cite{Chen_2017} the authors propose an iterative algorithm whereby a UAV seeks to maximise the throughput delivered to a specific user in a dense urban environment. The authors assume a dedicated BS backhaul with an interference-free LOS channel. The authors of \cite{Zeng_2016} and \cite{ZhangZeng_2017} tackle a similar scenario where a single user, UAV and BS are under consideration, with the UAV parameters optimised to maximise throughput and spectral efficiency, respectively. As before, the authors assume that the channels experience no interference from external sources.

\section{System Model}

\begin{figure}[t!]
\centering
	\includegraphics[width=.40\textwidth]{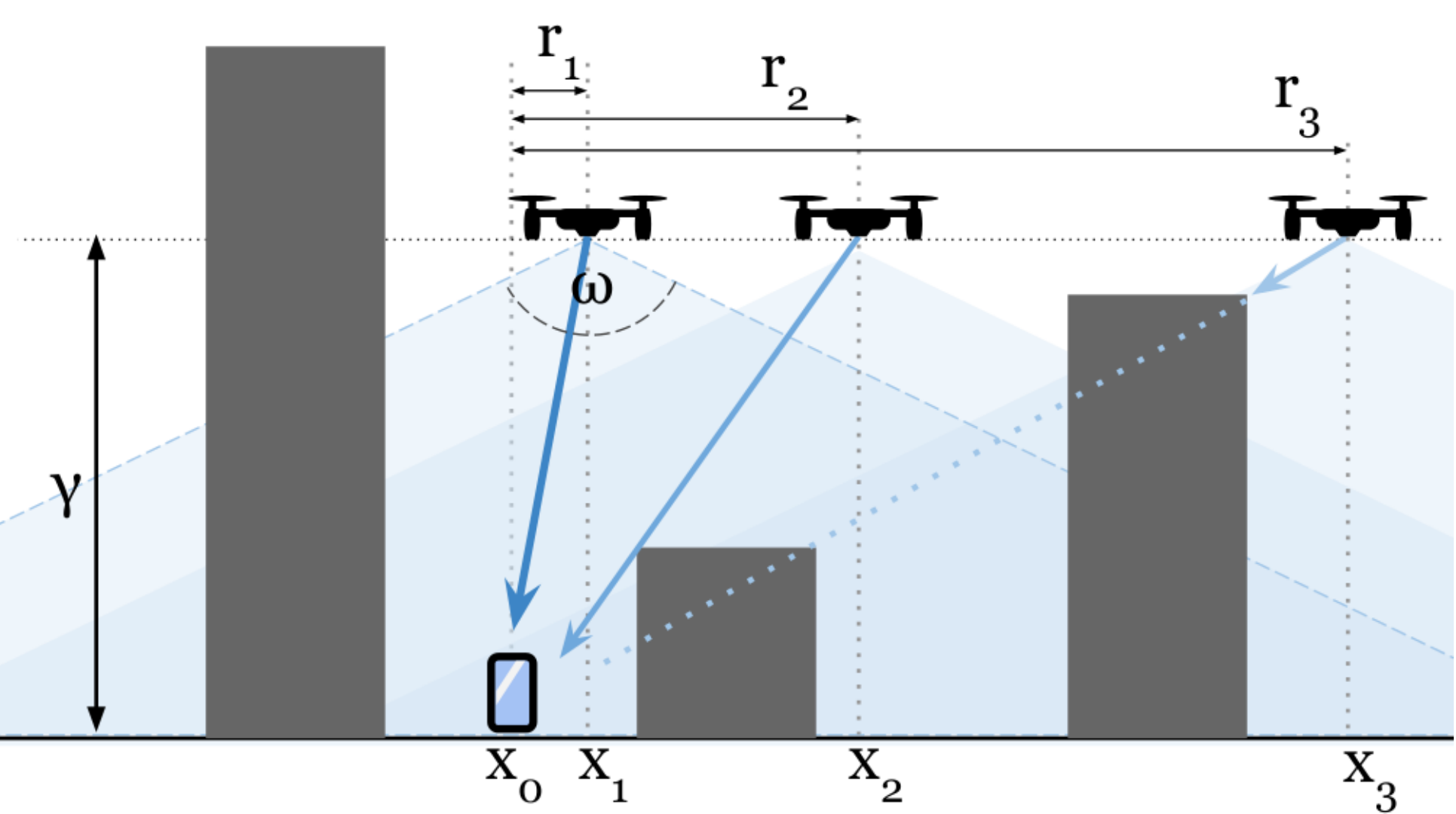}
	\caption{
	Side view showing the wireless link between the reference user and the UAVs. UAVs are at a height $\gamma$, with 2D coordinates $x_1,x_2$, $x_3$ and antenna beamwidth $\omega$. The user is serviced by the UAV with the strongest signal, while the remaining UAVs generate LOS and NLOS interference.
	\vspace{-5mm}
	}
	\label{fig:drone_network}
\end{figure}

\begin{figure}[t!]
\centering
	\includegraphics[width=.40\textwidth]{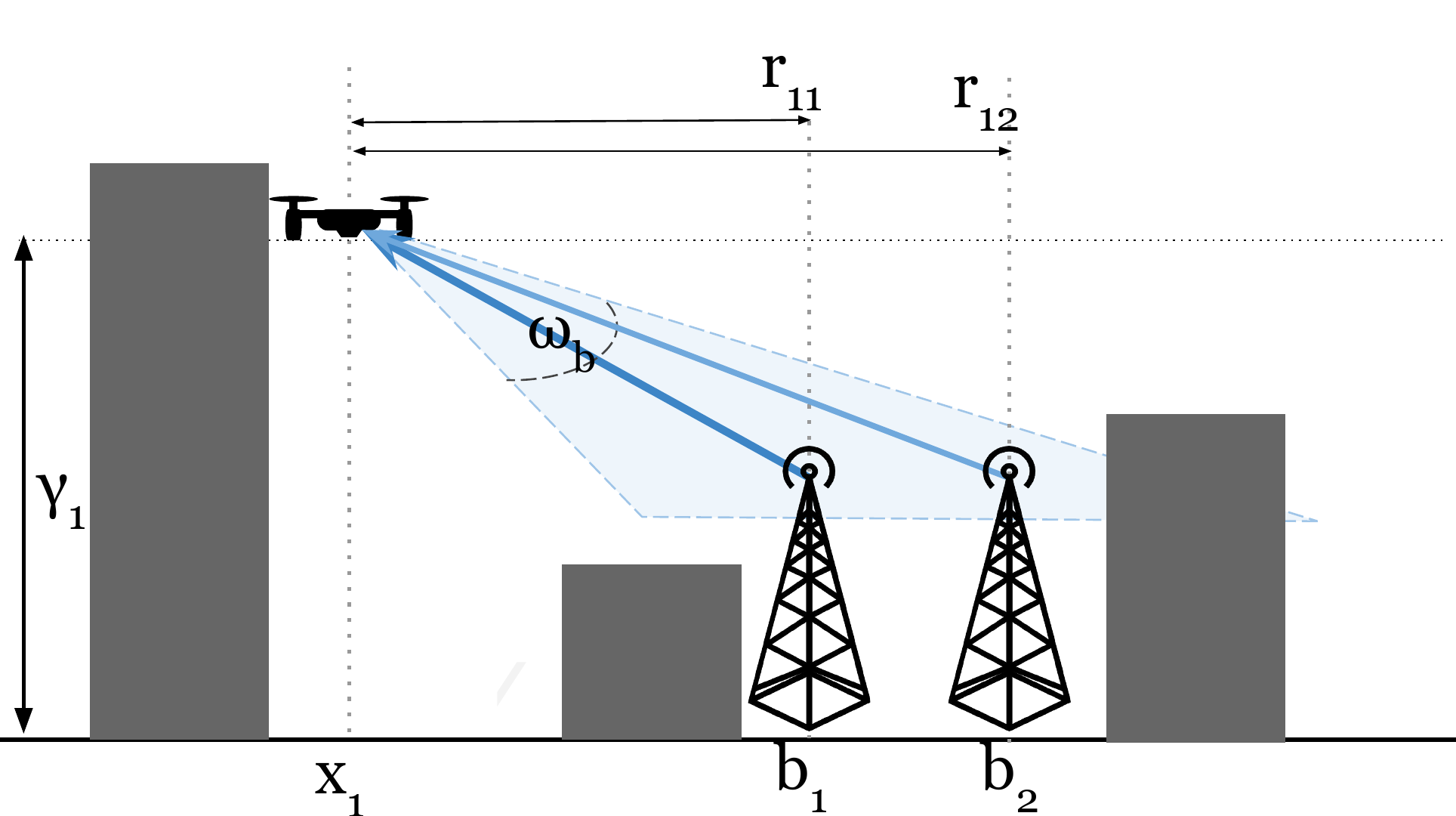}
	\caption{
	Side view showing the backhaul link of a UAV at $x_1$. The UAV aligns its backhaul antenna with beamwidth $\omega_B$ towards the nearest BS at $b_1$, which it selects as its backhaul BS. The BS at $b_2$ falls inside the main lobe of the UAV antenna and so creates interference for the signal.
	\vspace{-5mm}
	}
	\label{fig:drone_networkBH}
\end{figure}

Figures \ref{fig:drone_network} and \ref{fig:drone_networkBH} illustrate our system model. We consider a network of UAVs, a network of terrestrial BSs that provide wireless backhaul to the UAVs, and a reference user. In this paper we focus on downlink communication, which means we treat the UAVs as transmitters when communicating with the reference user and as receivers when communicating with the BSs. We position the reference user at $x_0\in\Rs$ and model the network of UAVs as a Poisson point process (PPP) $\Phi = \{x_1 , x_2 , ...\} \subset \Rs$ of intensity $\lambda$, where elements $x_i\in \Rs$ represent the projections of the UAV locations onto the $\Rs$ plane. The BS network is modelled as a PPP $\Phi_{B} = \{b_1 , b_2 , ...\} \subset \Rs$ of intensity $\lambda_{B}$. Given that a PPP is translation invariant with respect to the origin, we can set the 2D coordinates of our reference user to $x_0=(0,0)$. The reference user is assumed to be positioned on ground level with height $0$, and the UAVs and the BSs all have associated heights above ground: the BS height is denoted as $\gamma_{B}$ and is constant across the BS network; the heights of the individual UAVs are denoted as $\gamma_1,\gamma_2,... \subset \Rb$. UAVs have two directional antennas, one for communicating with the end user on the ground and another for communicating with the BS that is providing the backhaul. 

Each UAV selects the BS at the shortest distance to it as its source of wireless backhaul; the UAV then points its backhaul antenna towards the selected BS. If the backhaul signal received by the UAV from the selected BS does not meet a certain SINR threshold (defined in the next subsection) then the UAV is said to be in an outage state and does not provide wireless service to the end user; otherwise the UAV can service the end user using its end user antenna. The main beam of the end user antenna illuminates the area directly beneath the UAV, creating a cone of coverage. The reference user receives a signal from the UAVs which cast their coverage cone over it: the UAV which transmits the signal with the strongest received power is selected as the serving UAV and the signals from the remaining UAVs are treated as interference. The reference user is said to have coverage if the signal from the serving UAV meets an SINR threshold, otherwise the reference user is in outage. Each UAV's height is optimised to maximise the channel quality experienced by the reference user while ensuring that the UAV meets its own channel requirements for the backhaul.

The wireless channel between two devices will be affected by buildings in the environment, which form obstacles and break LOS links. We adopt the model in \cite{ITUR_2012}, which defines an urban environment as a collection of buildings arranged in a square grid. There are $\beta$ buildings per square kilometer, the fraction of area occupied by buildings to the total area is $\delta$, and each building has a height which is a Rayleigh-distributed random variable with scale parameter $\kappa$. The probability of a LOS between a transmitter-receiver pair is given in \cite{ITUR_2012} as 
\vspace{-1mm}
\begin{align}
&\Pd_{LOS}(\gamma_{tx},\gamma_{rx},r) = \nonumber \\
&\prod\limits_{n=0}^{\max(0,d-1)}\hspace{-1mm}\left(\hspace{-1mm}1-\exp\hspace{-1mm}\left(\hspace{-1mm}-\frac{\left(\max(
\gamma_{tx},\gamma_{rx}) - \frac{(n+1/2)|\gamma_{tx}-\gamma_{rx}|}{d}\right)\hspace{-1mm}^2}{2\kappa^2}\hspace{-1mm}\right)\hspace{-1mm}\right)
\label{eq:LOS}
\vspace{-1mm}
\end{align}

\noindent
where $\gamma_{tx}, \gamma_{rx},r$ denote the transmitter height, the receiver height and the horizontal distance between the transmitter and receiver, respectively, with $d = \floor*{r\sqrt{\beta\delta}}$. 

\subsection{End user link model}
The main beam of each UAV end user antenna illuminates the area directly beneath the UAV. The antenna has a circular radiation pattern with beamwidth $\omega$. Using the approximations (2-26) and (2-49) in \cite{Balanis_2005} and assuming perfect radiation efficiency, the antenna gain can be expressed as $\eta(\omega) = 16\pi/(\omega^2)$ inside the main lobe and 0 outside of the main lobe. It follows that the reference user will receive a signal from a UAV $i$ only if it falls inside the main lobe of the antenna, which will occur if the vertical angle between the reference user and the UAV $i$ $\phi_{i} = \tan^{-1}(\gamma_i/r_i)$ is greater than $\pi/2-\omega/2$, where $r_i = ||x_i||$ is the horizontal distance between the UAV $i$ and the reference user. We define the set of UAVs whose signals are received by the reference user as $\Phi_{\mathcal{W}} = \{x_i\in\Phi : \phi_{i} \geq \pi/2-\omega/2\}$. The probability of a LOS link between the UAV $i$ and the reference user is given as $\Pd_{LOS}(\gamma_i,0,r_i)$.

Let $S_{i}$ be the power received from the $i$th UAV by the reference user; for a given value of $r_i$ this is defined as $S_{i} = p\eta(\omega) H_{t_i} (r_i^2+\gamma_i^2)^{-\alpha_{t_i}/2}$ where $p$ is the UAV transmit power, $H_{t_i}$ is the random multipath fading component, and $\alpha_{t_i}$ is the pathloss exponent, where $t_i \in \{\text{L},\text{N}\}$ is an indicator variable which denotes whether the $i$th UAV has LOS or NLOS to the user. The SINR for the reference user can be described as $\sinr = S_{1}/(I_L + I_N+\sigma^2)$ where $S_1$ denotes the serving UAV signal, $I_L$ and $I_N$ denote the aggregate LOS and NLOS interference and $\sigma^2$ denotes the noise power.

The reference user is said to be covered by the UAV network if two conditions are met. First, there exists at least one UAV within $\Phi_{\mathcal{W}}$ such that the user is inside of an antenna main lobe from at least one of the UAVs, such that $|\Phi_{\mathcal{W}}|>0$, where $|.|$ denotes set cardinality. Second, the SINR experienced by the user is above some threshold $\theta$.

\subsection{Backhaul link model}
We denote the horizontal distance between a UAV $i$ and BS $j$ as $r_{ij} = ||x_i-b_j||$ and the vertical angle as $\phi_{ij} = \tan^{-1}((\gamma_i-\gamma_{B})/r_{ij})$. The BSs have identical characteristics, a transmit power $p_B$ and tri-sector antennas with downtilt angle $\phi_D$. The BS antenna downtilt reflects the fact that the terrestrial BSs have the primary purpose of servicing terrestrial users on the ground, with their antennas being configured accordingly. We model the BS antennas as being omnidirectional in the horizontal plane with horizontal antenna gain $\eta_{Bh}$. The BS vertical antenna gain between UAV $i$ and BS $j$ $\eta_{Bv}(\phi_{ij})$ is given by the 3GPP model \cite{3GPP_2010} as 

\begin{equation}
\eta_{Bv}(\phi_{ij}) = 10^{-\min\left(12\left(-\frac{\phi_{ij} + \phi_D}{10}\right)^2, 20 \right)/10},
\end{equation}

\noindent
and the total BS antenna gain $\eta_{B}(\phi_{ij})$ is given as 

\begin{equation}
\eta_{B}(\phi_{ij}) = \max\left(\eta_{Bh}\eta_{Bv}(\phi_{ij}),10^{-2.5}\right).
\end{equation}

\noindent
It can be seen that large values of $\phi_{ij}$ will correspond to low values of the BS antenna gain, reflecting the fact that the UAV will be receiving signals that radiate from BS antenna sidelobes. 

The UAV backhaul antenna is modelled as being directional, with a horizontal and vertical beamwidth $\omega_{B}$. The antenna has a rectangular radiation pattern with antenna gain $\eta(\omega_{B})$ inside the main lobe and 0 outside. If BS $j$ is the closest BS in $\Phi_{B}$ to UAV $i$ then the UAV $i$ selects it for its backhaul. The UAV orients itself to align its backhaul antenna towards BS $j$, and the antenna radiation pattern illuminates an area we denote as $\mathcal{W}_i \subset \Rs$. This area takes the shape of a ring sector centered on $x_i$ of arc angle equal to $\omega_{B}$ and major and minor radii $\Ru$ and $\Rl$, respectively, which are defined as 

\begin{align}
\Ru = 
\begin{cases}
\frac{\gamma_i - \gamma_{B}}{\tan(\phi_{ij}-\omega_{B}/2)} \hspace{-2mm} &\text{if} \hspace{3mm} \omega_{B}/2 < \phi_{ij} < \pi/2 - \omega_{B}/2 \\
\frac{\gamma_i - \gamma_{B}}{\tan(\pi/2 -\omega_{B})} \hspace{-2mm} &\text{if} \hspace{3mm} \phi_{ij} > \pi/2 - \omega_{B}/2 \\
\infty &\text{otherwise}
\end{cases}
\end{align}

\begin{align}
\Rl = 
\begin{cases}
\frac{\gamma_i - \gamma_{B}}{\tan(\phi_{ij}+\omega_{B}/2)} \hspace{2mm} &\text{if} \hspace{3mm} \phi_{ij} < \pi/2 - \omega_{B}/2  \\ 
0 &\text{otherwise}
\end{cases}
\end{align}

\noindent
Note that if $\omega_{B}\geq\pi/2$ major radius $\Ru$ will always have an infinite value. We denote the BSs inside of $\mathcal{W}_i$ as $\Phi_{BS_{\mathcal{W}_i}} \subset \Phi_{B}$. The probability of a LOS channel between UAV $i$ and BS $j$ is given as $\Pd_{LOS}(\gamma_{B},\gamma_i,r_{ij})$. The signal between UAV $i$ and BS $j$ is given as $p_B\eta(\omega_{B})\eta_{B}(\phi_{ij}) H_{t_{ij}} (r_{ij}^2+(\gamma_i-\gamma_{B})^2)^{-\alpha_{t_{ij}}/2}$, where $t_{ij}$ indicates whether UAV $i$ has LOS on BS $j$. The SINR for backhaul of the $i$th UAV can be described as $\sinr_i = S_{ij}/(I_L + I_N+\sigma^2)$ where $S_{ij}$ denotes the backhaul signal, and $I_L$ and $I_N$ denote the aggregate LOS and NLOS interference from the BSs in $\Phi_{BS_{\mathcal{W}_i}} \setminus b_j$. We define a SINR threshold $\theta_{B}$ for the UAV backhaul link: $\sinr_i<\theta_{B}$ represents the UAV $i$ failing to establish a backhaul of the required channel quality and therefore being in an outage state where it cannot serve any end user. 

\section{Mathematical Analysis}

In this section we provide analytical expressions for the coverage probability of the reference user from the network of UAVs, for the special case when all UAVs are positioned at the same altitude ($\gamma_i = \gamma$) and all UAVs have a backhaul. Because all of the UAVs have the same height they each provide coverage to a circular area of the same radius $\Rcone=\tan(\omega/2)\gamma$ centered below them. We define a bounded circular area $\mathcal{W}\subset \Rs$ of radius $\Rcone$ centered at the reference user: each UAV in $\Phi_{\mathcal{W}}$ whose signal can reach the reference user will have its 2D coordinates inside of $\mathcal{W}$. Note that $\Phi_{\mathcal{W}}$ is a PPP with the same intensity $\lambda$. As we are interested in the distances of the UAVs to the reference user, rather than their 2D coordinates in $\mathcal{W}$, we can apply the mapping theorem \cite{Haenggi_2013}[Theorem 2.34] to convert the 2D PPP $\Phi_{\mathcal{W}}$ into a 1D PPP $\Phi^\prime_{\mathcal{W}} \subset [0,\Rcone]$. $\Phi^\prime_{\mathcal{W}}$ is an inhomogeneous PPP with intensity function $\lambda'(r) = 2\pi\lambda r$ where the coordinates of the UAVs correspond to their horizontal distances to the user. Note that we drop the index $i$ as the horizontal distances of the UAVs in $\Phi^\prime_{\mathcal{W}}$ have the same distribution irrespective of their index values. We partition $\Phi^\prime_{\mathcal{W}}$ into two PPPs which contain the LOS and NLOS UAVs, denoted as $\Phi^\prime_{\mathcal{W}L} \subset [0,\Rcone]$ and $\Phi^\prime_{\mathcal{W}N} \subset [0,\Rcone]$, respectively, with intensity functions $\lambda'_{L}(r) = \Pd_{LOS}(r) 2\pi\lambda r$ and $\lambda'_{N}(r) = (1-\Pd_{LOS}(r)) 2\pi\lambda r$, where for notational simplicity we use $\Pd_{LOS}(r)$ to denote $\Pd_{LOS}(\gamma,0,r)$, the LOS probability function \Eq{LOS}. In effect, the LOS probability function acts as a thinning function \cite{Haenggi_2013} which removes (thins) UAVs from the PPP $\Phi^\prime_{\mathcal{W}}$ with probability $(1-\Pd_{LOS}(r))$ to form $\Phi^\prime_{\mathcal{W}L}$ from the remaining UAVs and $\Phi^\prime_{\mathcal{W}N}$ from those that are thinned.

\subsection{Distribution of the distance to the serving UAV}
The received signal power is affected by the distance between the UAV and the reference user, the multipath fading $H_{t_i}$ and the pathloss $\alpha_{t_i}$. Note that $\alpha_N > \alpha_L$ to represent the attentuation that happens when a wireless signal encounters obstacles. As a result of this difference in pathloss, UAVs which are physically closer to the reference user but that are blocked by buildings will have lower received signal strength than UAVs which are further away but have LOS to the user. This introduces a complication to determining the serving UAV for the reference user. In the literature, when UAVs are assumed to all have identically performing wireless channels to the user the UAV that is closest to the user will also be the UAV with the strongest received signal power; therefore, the user will always be serviced by the closest UAV, and the UAVs beyond the serving UAV distance act as interferers. To account for the LOS blocking effects we adopt a different approach to determining which UAV a user associates with. The user will be serviced by the UAV that provides the strongest received signal power. If the multipath fading effect $H_{t_i}$ is averaged out the strongest signal power will come from \textit{either} the closest LOS UAV to the user \textit{or} the closest NLOS UAV. We denote the horizontal distance to the serving UAV as the random variable $R_1$ and refer to it as the distance to the serving UAV. 

\textit{Proposition 1:} The probability density functions for the horizontal distance to the serving UAV are given as 

  \vspace{-2mm} 
  \begin{align}
  & f_{R_{1},t_1}(r_1,t_1 = \text{L}) = \Pd_{LOS}(r_1) 2\pi\lambda r_1 \nonumber \\
  &\cdot\exp\left(-2 \pi \lambda \int\limits_{0}^{r_1}\Pd_{LOS}(r)r\dr r\right) \nonumber \\ &\cdot\exp\left(-2\pi\lambda\int\limits_{0}^{c_N}\big(1-\Pd_{LOS}(r)\big)r\dr r\right),
  \label{eq:pdfLOS}
  \end{align}
  \vspace{-2mm} 
  
  \begin{align}
  &f_{R_{1},t_1}(r_1,t_1 = \text{N}) = \big(1-\Pd_{LOS}(r_1)\big) 2\pi\lambda r_1 \nonumber \\
  &\cdot\exp\left(-2 \pi \lambda \int\limits_{0}^{r_1}\big(1-\Pd_{LOS}(r)\big)r\dr r\right) \nonumber \\ &\cdot \exp\left(-2\pi\lambda\int\limits_{0}^{c_L}\Pd_{LOS}(r)r\dr r\right),
  \label{eq:pdfNLOS}
  \end{align}

for the cases when the serving UAV has a LOS and an NLOS to the user, respectively, with 

\begin{equation}
c_N = \sqrt{\max(0,(r_1^2+\gamma^2)^{\alpha_L/\alpha_N} - \gamma^2)}
\end{equation}

and

\begin{equation}
c_L = \min(\Rcone,\sqrt{(r_1^2+\gamma^2)^{\alpha_N/\alpha_L} - \gamma^2}).
\end{equation}

\textit{Proof: } If a user is serviced by an LOS UAV a horizontal distance $r_1$ away then this means that there are no LOS UAVs with horizontal distance smaller than $r_1$ to the user and that there are no NLOS UAVs with horizontal distance smaller than some lower distance bound $c_N$, where $p\eta(\omega)(r_1^2+\gamma^2)^{-\alpha_{L}/2} = p\eta(\omega)(c_N^2+\gamma^2)^{-\alpha_{N}/2}$. We can derive the expression for the probability distribution of the serving UAV distance $R_1$ when the serving UAV is within LOS by combining the probability that the closest LOS UAV in the PPP $\Phi^\prime_{\mathcal{W}L}$ is at $r_1$ (as given in \cite{Haenggi_2013}) with the probability that no NLOS UAV exists within a distance $c_N$. The probability distribution for the distance to the serving UAV when it has NLOS to the user can be obtained following the same logic as above. 

\textit{Remark 1: } If $c_N=0$ this means that the LOS serving UAV is close enough to the user that no NLOS UAV will be able to provide a stronger signal no matter how close to the user. If $c_L = \Rcone$ this denotes that all LOS UAVs must be outside of the window $\mathcal{W}$ for the user to receive the strongest signal from an NLOS UAV at a distance $r_1$. 

\textit{Proposition 2:} Having obtained the probability distributions of the serving UAV distance we can calculate the probability that a user will have an LOS channel to the UAV that is serving it as  
 
 \begin{equation}
 \Pd(t_1 = \text{L}||\Phi_{\mathcal{W}}|>0) = \frac{\int\limits_{0}^{\Rcone}f_{R_{1},t_1}(r_1,t_1 = \text{L})\dr r_1}{\Pd(|\Phi_{\mathcal{W}}|>0)},
 \label{losServ}
 \end{equation}
 
 where
 
 \begin{equation}
 \Pd(|\Phi_{\mathcal{W}}|>0) = 1- \exp(-\pi\lambda\Rcone^2).
 \label{eq:assProb}
 \end{equation}
 
 \textit{Proof: } The probability that the user has LOS to its serving UAV is the probability of the LOS serving UAV being within range of the user conditioned on there existing at least one UAV within range of the user, that is, inside the area $\mathcal{W}$.

\subsection{Aggregate LOS \& NLOS interference}

Having obtained the lower bounds on the horizontal distances at which the LOS and NLOS interfering UAVs may be found, we can now characterise the expressions for the aggregate LOS and NLOS interference. The LOS and NLOS interferers will belong to the sets $\Phi^\prime_{\mathcal{W}L} \setminus [0,c_L(t_1)]$ and $\Phi^\prime_{\mathcal{W}N} \setminus [0,c_N(t_1)]$, where $c_L(t_1)$ and $c_N(t_1)$ denote the lower bounds on LOS and NLOS interferer distances, as functions of the serving UAV type. If the serving UAV is within LOS then $c_L(\text{L}) = r_1$ and $c_N(\text{L}) = \sqrt{\max(0,(r_1^2+\gamma^2)^{\alpha_L/\alpha_N} - \gamma^2)}$, and if the serving UAV is NLOS then $c_L(\text{N}) =  \min(\Rcone,\sqrt{(r_1^2+\gamma^2)^{\alpha_N/\alpha_L} - \gamma^2})$ and $c_N(\text{N}) = r_1$. The aggregate LOS and NLOS interference is then described as $I_L=\sum_{r\in\Phi^\prime_{\mathcal{W}L}\setminus [0,c_L(t_1)]}p\eta(\omega) H_L (r^2+\gamma^2)^{-\alpha_L/2}$ and $I_N=\sum_{r\in\Phi^\prime_{\mathcal{W}N}\setminus [0,c_N(t_1)]}p\eta(\omega) H_N (r^2+\gamma^2)^{-\alpha_N/2}$. 


\subsection{Conditional coverage probability}
\noindent
Deriving an expression for the coverage probability involves the intermediate steps of deriving an expression for the conditional coverage probability in terms of the Laplace transforms of the interference produced by LOS and NLOS interferers, followed by deriving analytical expressions for these Laplace transforms. The conditional coverage probability is defined as the probability that the SINR of the downlink signal from the serving UAV to the user is above a threshold $\theta$, given $R_1=r_1$. 

\textit{Proposition 3: } The conditional coverage probability for an LOS serving UAV $\Pd(\sinr\geq \theta |R_1=r_1,t_1 = \text{L})$ is given as 

\begin{align}
&\sum\limits_{n=0}^{m_L-1}\frac{s_L^n}{n!} (-1)^n 
 \cdot \sum_{i_L+i_N+i_{\sigma}=n}\frac{n!}{i_L!i_N!i_{\sigma}!} \nonumber \\
 &\cdot(-(p\eta(\omega))^{-1}\sigma^2)^{i_{\sigma}}\exp(-(p\eta(\omega))^{-1}s_L\sigma^2) \nonumber \\
 &\cdot\frac{d^{i_L} \Lc_{I_L}((p\eta(\omega))^{-1}s_L)}{ds_L^{i_L}} \frac{d^{i_N}\Lc_{I_N}((p\eta(\omega))^{-1}s_L)}{ds_L^{i_N}},
\label{eq:condProb3}
\end{align}

\noindent
where $s_{L}= m_{L}\theta(r_1^2+\gamma^2)^{\alpha_{L}/2}$, $m_{L}$ is the Nakagami-$m$ fading term for a LOS channel, $\Lc_{I_L}$ and $\Lc_{I_N}$ are the Laplace transforms of the aggregate LOS and NLOS interference, respectively, and the second sum is over all the combinations of non-negative integers $i_L,i_N$ and $i_{\sigma}$ that add up to $n$. The conditional coverage probability given an NLOS serving UAV $\Pd(\sinr\geq \theta |R_1=r_1,t_1 = \text{N})$ is calculated as in \Eq{condProb3} with $m_N$, $\alpha_N$ and $s_N$ replacing $m_L$, $\alpha_L$ and $s_L$.

\textit{Proof: } Considering Nakagami-$m$ fading, the conditional coverage probability $\Pd(\sinr\geq \theta |R_1=r_1)$ is obtained following (21) in \cite{Chetlur_2017} as

\begin{equation}
 \sum\limits_{n=0}^{m_{t_1}-1}\frac{s_{t_1}^n}{n!} (-1)^n \frac{d^n \Lc_{I}((p\eta(\omega))^{-1}s_{t_1})}{ds_{t_1}^n},
\end{equation}

\noindent
where $\Lc_{I}$ denotes the Laplace transform of the total interference. The LOS and NLOS interferers are distributed independently of one another; the proof of this is similar to the proof in \cite{Haenggi_2013} and is omitted here. Due to this, the Laplace transform above can be separated into a product of the Laplace transforms of the aggregate LOS and aggregate NLOS interference, along with the introduction of the noise-related term.

\subsection{Laplace transform of aggregate interference}

\textit{Proposition 4: } The Laplace transform of the aggregate LOS interference given an LOS serving UAV is expressed as 

\begin{align}
&\Lc_{I_L}((p\eta(\omega))^{-1}s_L) = \nonumber \\
&=\exp\Bigg(-\pi\lambda\sum\limits_{j=\floor*{c_L(\text{L})\sqrt{\beta\delta}}}^{\floor*{\Rcone\sqrt{\beta\delta}}} \Pd_{LOS}(l) \sum\limits_{k=1}^{m_L}\binom{m_L}{k}(-1)^{k+1}\nonumber \\
&\cdot\bigg((\Upper^2+\gamma^2)\mbox{$_2$F$_1$}\Big(k,\frac{2}{\alpha_L};1+\frac{2}{\alpha_L};  -\frac{m_L(\Upper^2+\gamma^2)^{\alpha_L/2}}{s_L}\Big) \nonumber\\
&-(\Lower^2+\gamma^2)\mbox{$_2$F$_1$}\Big(k,\frac{2}{\alpha_L};1+\frac{2}{\alpha_L};-\frac{m_L(\Lower^2+\gamma^2)^{\alpha_L/2}}{s_L}\Big)\bigg)\Bigg) 
\label{eq:laplaceFinal}
\end{align}

where $\Lower = \max(c_L(\text{L}),j/\sqrt{\beta\delta})$ and $\Upper = \min(\Rcone,(j+1)/\sqrt{\beta\delta})$ , and $\mbox{$_2$F$_1$}(a,b;c;z)$ denotes the Gauss hypergeometric function.


\textit{Remark 2: } The Laplace transform for the NLOS interferers $\Lc_{I_N}((p\eta(\omega))^{-1}s_L)$ is obtained by simply substituting $\Pd_{LOS}(l$) with $(1-\Pd_{LOS}(l))$, $c_L(\text{L})$ with $c_N(\text{L})$, $m_L$ with $m_N$ and $\alpha_L$ with $\alpha_N$. The above integration is for the case when the serving UAV is LOS; if the serving UAV is NLOS we substitute $s_L$ with $s_N$ and $c_L(\text{L})$ with $c_L(\text{N})$. The higher derivatives of the Laplace transforms become cumbersome to solve manually for larger values of the serving UAV fading parameter, so in order to obtain an analytical expression we treat the Laplace transforms as composite functions and apply Fa\`{a} di Bruno's formula for higher derivatives. This is given in the Appendix.

\textit{Proof: } The Laplace transform of the aggregate LOS interference $\Lc_{I_L}((p\eta(\omega))^{-1}s_L)$ given an LOS serving UAV is expressed as
\begin{align}
&\Ed_{\Phi^\prime_{\mathcal{W}L}}\bigg[\prod_{r\in\Phi^\prime_{\mathcal{W}L} \setminus [0,c_L(\text{L})]}\hspace{-3mm}\Ed_{H_L} \left[\exp\Big(-s_L H_L (r^2+\gamma^2)^{-\alpha_L/2}\Big)\right]\bigg]  \nonumber \\
&\overset{(a)}{=}\Ed_{\Phi^\prime_{\mathcal{W}L}}\left[\prod_{r\in\Phi^\prime_{\mathcal{W}L} \setminus [0,c_L(\text{L})]}g(r,s_L,m_L,\alpha_L)\right]  \nonumber \\
&\overset{(b)}{=}\exp\Bigg(-\int\limits_{c_L(\text{L})}^{\Rcone} \left(1-g(r,s_L,m_L,\alpha_L)\right) \lambda'_{L}(r) \dr r\Bigg) 
\label{eq:laplace}
\end{align}

\noindent
where 
\begin{equation}
 g(r,s_L,m_L,\alpha_L) = \left(\frac{m_L}{s_L(r^2+\gamma^2)^{-\alpha_L/2} + m_L}\right)^{m_L}   \nonumber,
\end{equation}

\noindent
 $(a)$ comes from Nakagami-$m$ fading having a gamma distribution, $(b)$ comes from the probability generating functional of the PPP \cite{Haenggi_2013} and $\lambda'_{L}(r) = \Pd_{LOS}(r) 2\pi\lambda r$. From the definition of the LOS probability function we can observe that $\Pd_{LOS}(r)$ is a step function. We use this fact to separate the integral above into a sum of weighted integrals, resulting in the following expression:
 
 \begin{equation}
2\pi\lambda\sum\limits_{j=\floor*{c_L(\text{L})\sqrt{\beta\delta}}}^{\floor*{\Rcone\sqrt{\beta\delta}}} \Pd_{LOS}(l)\int\limits_{l}^{u}(1-g(r,s_L,m_L,\alpha_L))r\dr r 
\label{eq:laplaceSum}
 \end{equation}

\noindent
The integral $\int\limits_{\Lower}^{\Upper}(1-g(r,s_L,m_L,\alpha_L))r\dr r$ can then be expressed as 

\begin{align}
&\overset{(a)}{=} \int\limits_{(\Lower^2+\gamma^2)^{1/2}}^{(\Upper^2+\gamma^2)^{1/2}}\left(1-\left(\frac{m_L}{s_Ly^{-\alpha_L} + m_L}\right)^{m_L}\right)y\dr y \nonumber \\
& \overset{(b)}{=} \frac{ 1}{\alpha_L}\int\limits_{(\Lower^2+\gamma^2)^{\alpha_L/2}}^{(\Upper^2+\gamma^2)^{\alpha_L/2}}\left(1-\left(1-\frac{1}{1 + zm_Ls_L^{-1}}\right)^{m_L}\right)z^{2/\alpha_L - 1}\dr z \nonumber \\
& \overset{(c)}{=} \frac{    1}{\alpha_L}\sum\limits_{k=1}^{m_L}\binom{m_L}{k}(-1)^{k+1}\int\limits_{(\Lower^2+\gamma^2)^{\alpha_L/2}}^{(\Upper^2+\gamma^2)^{\alpha_L/2}}\frac{z^{2/\alpha_L - 1}}{(1+zm_Ls_L^{-1})^k}\dr z ,\nonumber \\
&\overset{(d)}{=} \frac{1}{2}\sum\limits_{k=1}^{m_L}\binom{m_L}{k}(-1)^{k+1}\nonumber \\
&\cdot\bigg((\Upper^2+\gamma^2)\mbox{$_2$F$_1$}\Big(k,\frac{2}{\alpha_L};1+\frac{2}{\alpha_L};  -\frac{m_L(\Upper^2+\gamma^2)^{\alpha_L/2}}{s_L}\Big) \nonumber\\
&-(\Lower^2+\gamma^2)\mbox{$_2$F$_1$}\Big(k,\frac{2}{\alpha_L};1+\frac{2}{\alpha_L};-\frac{m_L(\Lower^2+\gamma^2)^{\alpha_L/2}}{s_L}\Big)\bigg),
\label{eq:laplace_final}
\end{align}

\noindent
where $(a)$ stems from the substitution $y=(r^2+\gamma^2)^{1/2}$, $(b)$ from the substitution $z = y^{a_L}$, $(c)$ from applying binomial expansion and $(d)$ from using \cite{Ryzhik_2007}[Eq. 3.194.1]. Inserting this solution into \Eq{laplaceSum} we obtain an expression for the Laplace transform of the LOS interferers \Eq{laplaceFinal}.

\subsection{Coverage probability}

In this subsection we present the main result of our mathematical analysis.

\textit{Theorem 1 :} The coverage probability of a reference user being served by a network of PPP-distributed UAVs equipped with directional antennas a certain height above ground is given as 

\begin{multline}
\Pd(\sinr\geq\theta) = \\
\int\limits_{0}^{\Rcone}\bigg(\Pd(\sinr\geq \theta |R_1=r_1,t_1 = \text{L})f_{R_{1},t_1}(r_1,t_1 = \text{L}) \\
+\Pd(\sinr\geq \theta |R_1=r_1,t_1 = \text{N})f_{R_{1},t_1}(r_1,t_1 = \text{N})\bigg) \dr r_1 ,
 \label{eq:pcov_final}
\end{multline}

\noindent
where $f_{R_{1},t_1}(r_1,t_1 = \text{L})$ and $f_{R_{1},t_1}(r_1,t_1 = \text{N})$ are derived in Proposition 1, $\Pd(\sinr\geq \theta |R_1=r_1,t_1 = \text{L})$ and $\Pd(\sinr\geq \theta |R_1=r_1,t_1 = \text{N})$ are derived in Propositions 3 and 4. 

\textit{Proof: }To obtain the overall coverage probability for the reference user in the network we decondition the conditional coverage probability with respect to the indicator variable $t_1$ by multiplying by the two probability distributions given in \Eq{pdfLOS} and \Eq{pdfNLOS}. We then decondition with respect to the horizontal distance random variable $R_1$ via integration.

\section{Numerical Results}
\begin{figure}[t!]

\centering
	\includegraphics[width=.45\textwidth]{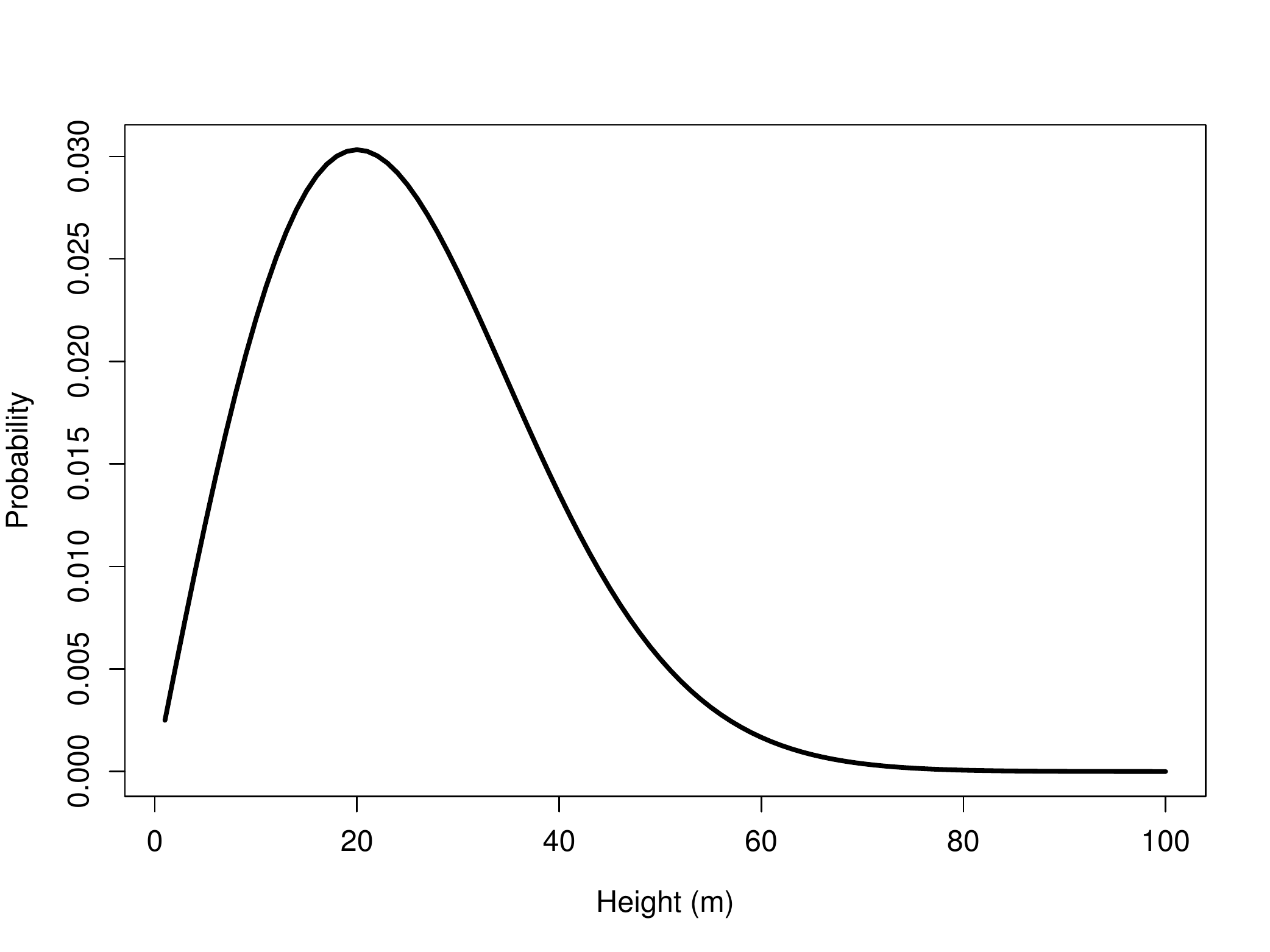}
    \vspace{-5mm}
	\caption{
	Probability density function of a typical building height in the urban environment.
	}
	\label{fig:buildDist}
	\vspace{-3mm}
\end{figure}

In this section we demonstrate how our model can provide insight into the behaviour of low-altitude UAV networks in urban environments. In the first subsection we validate our mathematical results against simulations for the special case where all UAVs operate at the same height and are assumed to have guaranteed backhauls. In the second subsection we analyse the backhaul between a typical UAV and the BS network for different parameters to analyse the probability of the UAV being able to establish a backhaul. In the third subsection we simulate a scenario where UAVs have to serve the end users while simultaneously meeting their backhaul requirements. Unless stated otherwise the parameters used for the numerical results are from Table \ref{tab:table}. \Fig{buildDist} shows the pdf of the height of a typical building in our environment.

\begin{table}[b!]
\vspace{-3mm}
\begin{center}
\caption{Numerical Result Parameters}
\begin{tabular}{ |c|c| } 
 \hline
 Parameter & Value  \\ 
 \hline
 $\omega$ & \unit[$150$]{deg} \\
 $\omega_B$ & \unit[$20$]{deg} \\
 $\alpha_L$ & 2.1 \\
 $\alpha_N$ & 4 \\
 $m_L$ & 3 \\
 $m_N$ & 1 \\
 $p$ & \unit[0.1]{W} \\
 $p_B$ & \unit[40]{W} \\
 $\eta_{Bh}$ & 0.31 \\
 $\phi_D$ & \unit[$10$]{deg} \\
 $\theta$ & \unit[0]{dB} \\
 $\theta_{B}$ & \unit[10]{dB} \\
 $\lambda_{B}$ & \unit[5]{$/\text{km}^2$} \\
 $\gamma_{B}$ &  \unit[30]{m} \\
 $\sigma^2$ & \unit[$10^{-9}$]{W} \\
 $\beta$ & \unit[300]{$/\text{km}^2$}\\
 $\delta$ & 0.5\\
 $\kappa$ & \unit[20]{m} \\
 \hline
\end{tabular}
 \label{tab:table}
\end{center}
\end{table}

\subsection{Coverage probability under guaranteed backhaul}
\begin{figure}[t!]

\centering
	\includegraphics[width=.45\textwidth]{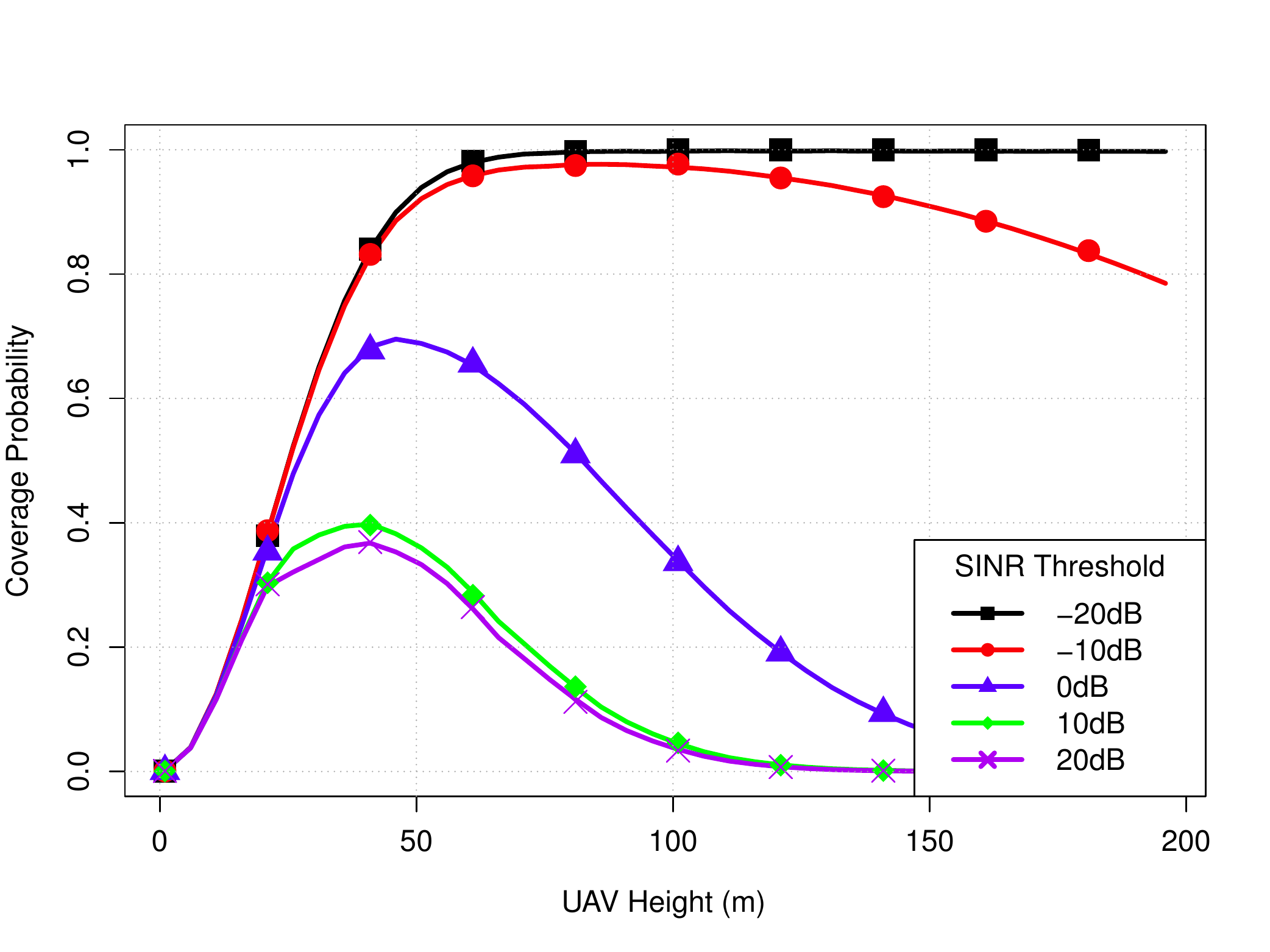}
    \vspace{-5mm}
	\caption{
	Coverage probability given a UAV density of \unit[25]{$\text{/km}^2$}.
	}
	\label{fig:LOSSINR}
	\vspace{-3mm}
\end{figure}

\begin{figure}[t!]
\centering
	\includegraphics[width=.45\textwidth]{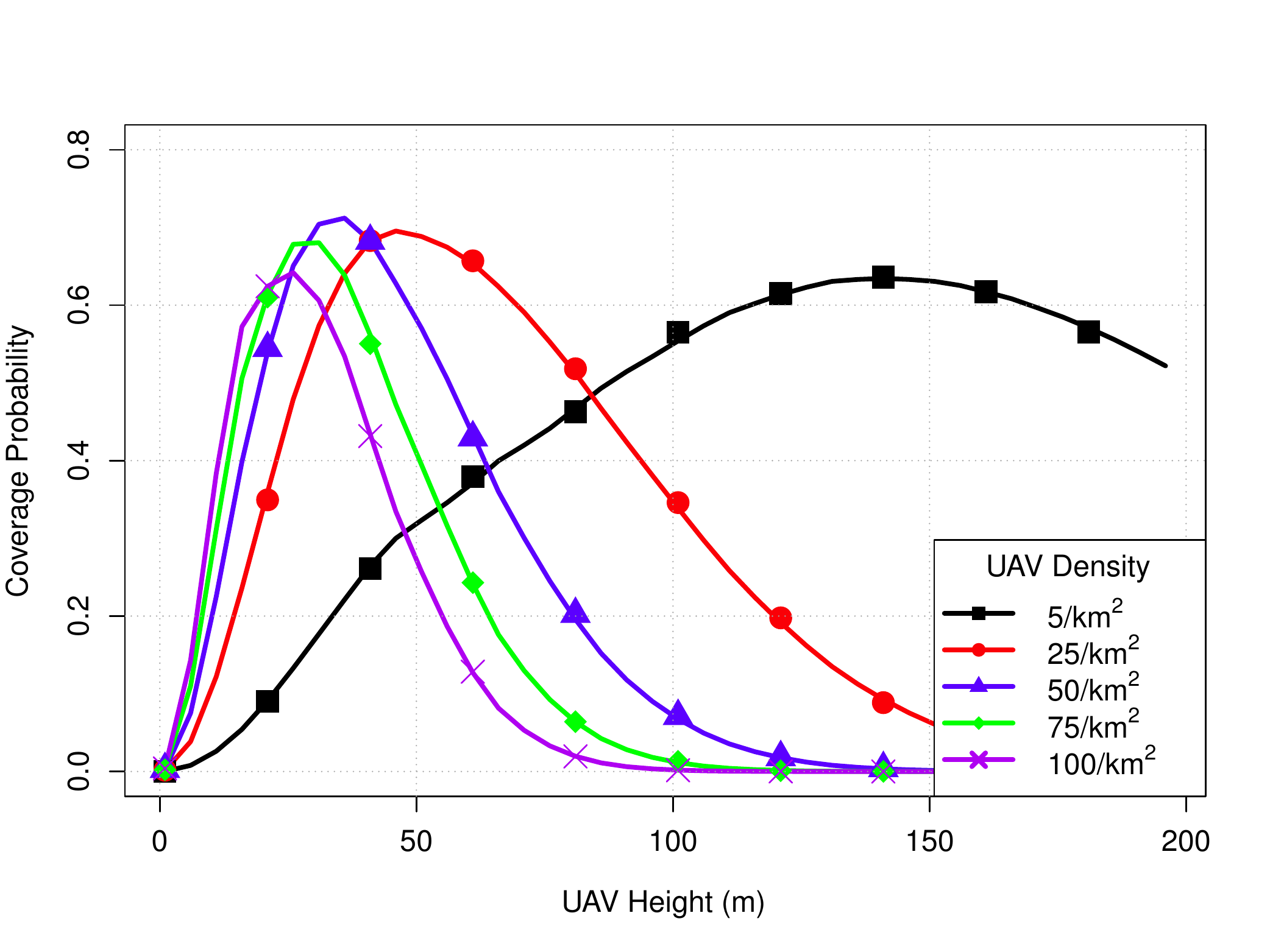}
	\caption{
	Coverage probability for multiple UAV densities, given an SINR threshold of \unit[0]{dB}.
	}
	\label{fig:LOSDensity}
	\vspace{-5mm}
\end{figure}

\begin{figure}[t!]
\centering
	\includegraphics[width=.45\textwidth]{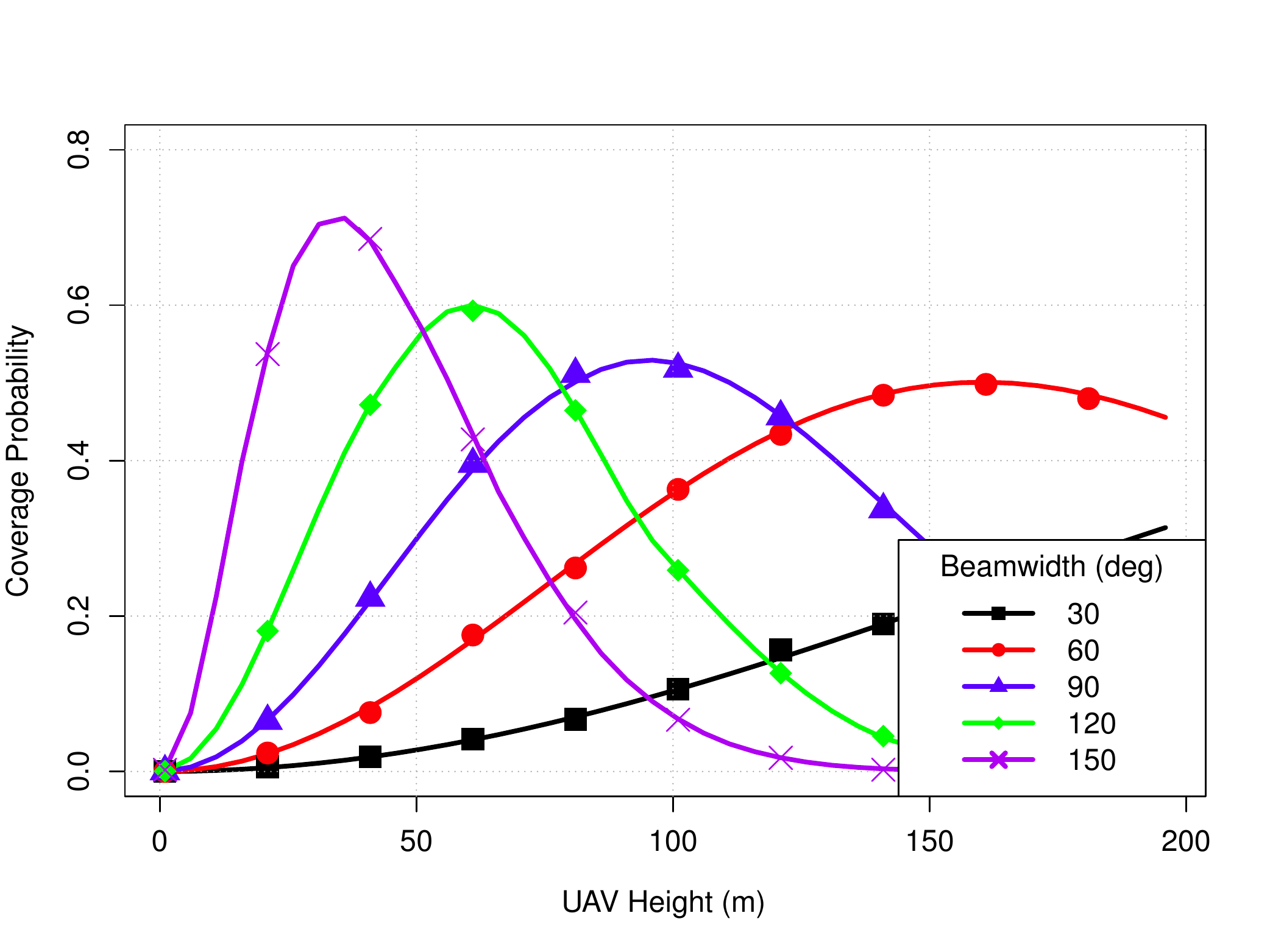}
	\caption{
	Coverage probability given an SINR threshold of \unit[0]{dB} and UAV density of \unit[50]{$\text{/km}^2$}.
	}
	\label{fig:LOSBeamwidth}
\end{figure}

\begin{figure}[t!]
\centering
	\includegraphics[width=.45\textwidth]{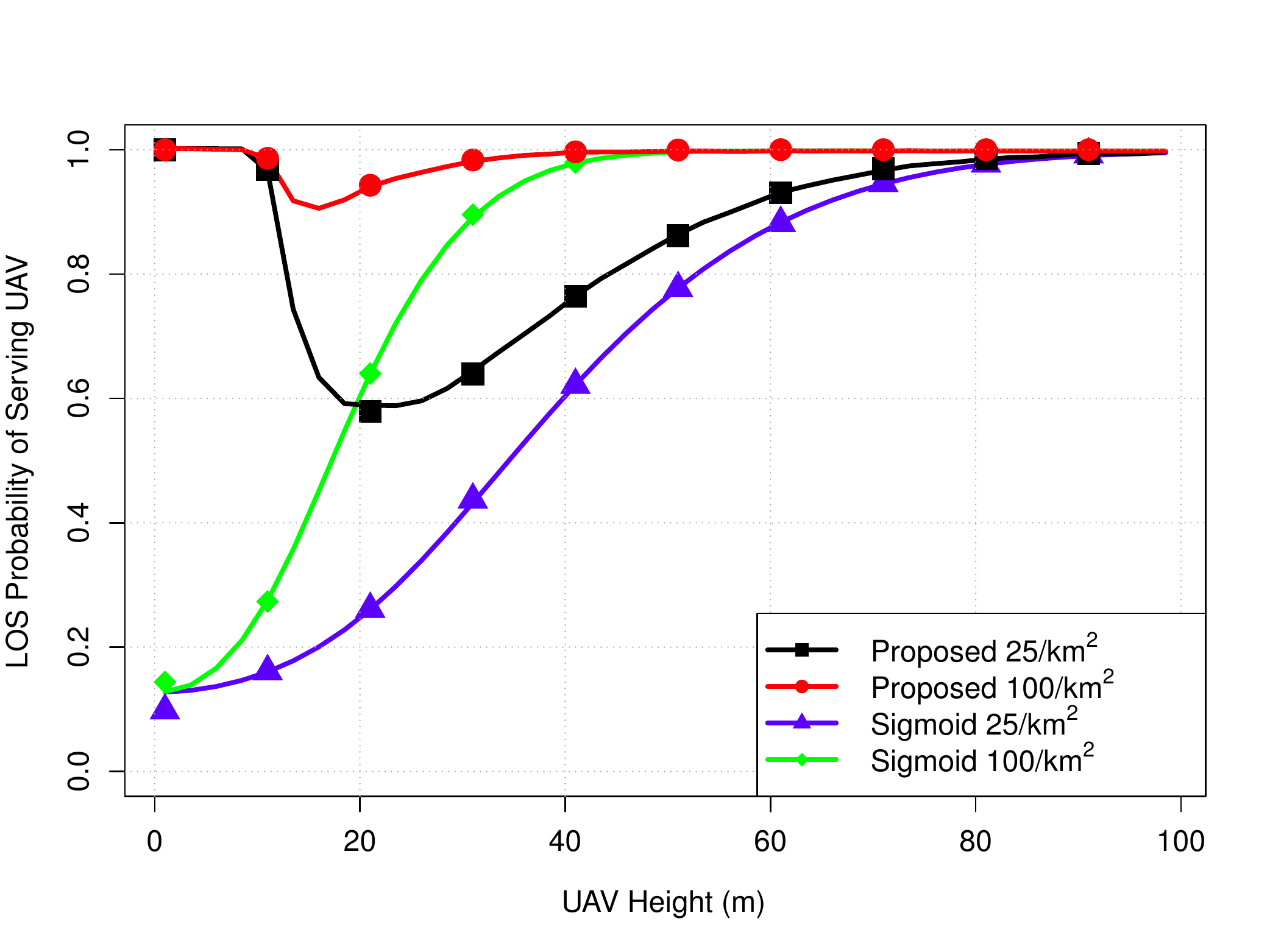}
	\caption{
	Probability of an LOS channel between a user and its serving UAV.
	}
	\label{fig:Comparison}
\end{figure}

We generate random deployments of UAVs across multiple Monte Carlo (MC) trials and record the mean coverage probability values, given the assumption that each UAV has a guaranteed backhaul into the core network. In Figures \ref{fig:LOSSINR} to \ref{fig:Comparison}, solid lines denote the analytical values for the coverage probability (from \Eq{pcov_final}) and the markers denote results from MC trials. 

\Fig{LOSSINR} shows the effect of the UAV height on the coverage probability, given different SINR threshold values. We can see that initially the coverage probabilities for all the SINR thresholds improve as we increase the height. This is due to the UAVs increasing their coverage areas, which maximises the probability that there is at least one UAV within range of the user and providing sufficient SINR. Past a certain height, however, coverage probability starts to decrease with increased height. This shows how the signals are more vulnerable to the increasing number of LOS interferers that appear as the UAV heights increase. 

In \Fig{LOSDensity} we consider how increasing the UAV height affects the coverage probability, for UAV networks of different densities. The figure shows how for each UAV density there is a corresponding height which maximises the coverage probability, and that this height decreases as the density increases. This is explained by considering the effect of the buildings on the networks. At low densities the serving UAV for the reference user may be concealed behind several buildings, and increasing the UAV height increases the chances of establishing an LOS channel. The low number of interferers within range means that as the channel between a user and its serving UAV improves, the net impact on the network performance is positive. In a high density network the serving UAV to a user is likely to be close enough that there are few buildings to interfere with the signal. The buildings in this scenario do not impede the serving UAV signal but instead shield the user from interfering UAVs a further distance away. Increasing the UAV height then will expose the user to these interferers while at the same time worsening the serving signal, resulting in a drop in coverage.

In \Fig{LOSBeamwidth} we demonstrate the effect of the UAV antenna beamwidth on the coverage probability. The  coverage curves suggest that narrower beamwidths perform best at larger UAV heights. This is due to the effect of the beawmidths on the probability of the user being within range of a UAV: narrow beamwidth UAVs create a narrow coverage cone and as a result must operate at larger heights to ensure that users can be within range of service. We see that each beamwidth value has an associated optimum UAV height for a given SINR threshold, UAV density and building environment.

In \Fig{Comparison} we show the probability that a user that is within range of the network will have an LOS link to the serving UAV under our LOS model and the sigmoid approximation adopted in \cite{AlHourani_2014}, given two different UAV densities. The sigmoid model gives the LOS probability as a function of the vertical angle between the UAV and the user, and as such when the UAV is close to the ground the LOS probability to its users approaches zero, despite the fact that the UAV coverage cone is very small and therefore the users are very close to the UAV. As the height increases this probability steadily improves due to the increasing vertical angle. Our model captures a more realistic behaviour of the LOS probability; when the UAV is very low to the ground, due to the size of its coverage cone its users are close enough that no LOS-blocking buildings are in the way, ensuring an LOS probability approaching unity. As the height increases the increasing coverage cone allows users further away to associate to the UAV, resulting in more users behind buildings being served by the UAV, which negatively affects the LOS probability. Finally, as the UAV ascends above the majority of buildings this LOS probability steadily improves to reflect the fact that there will be fewer buildings tall enough to block the UAV-user link.

\subsection{UAV backhaul} 
In this subsection we explore the ability of a typical UAV to establish a wireless backhaul above a certain SINR threshold through a terrestrial BS. The results are generated using multiple MC trials. 
\begin{figure}[b!]
\centering
	\includegraphics[width=.45\textwidth]{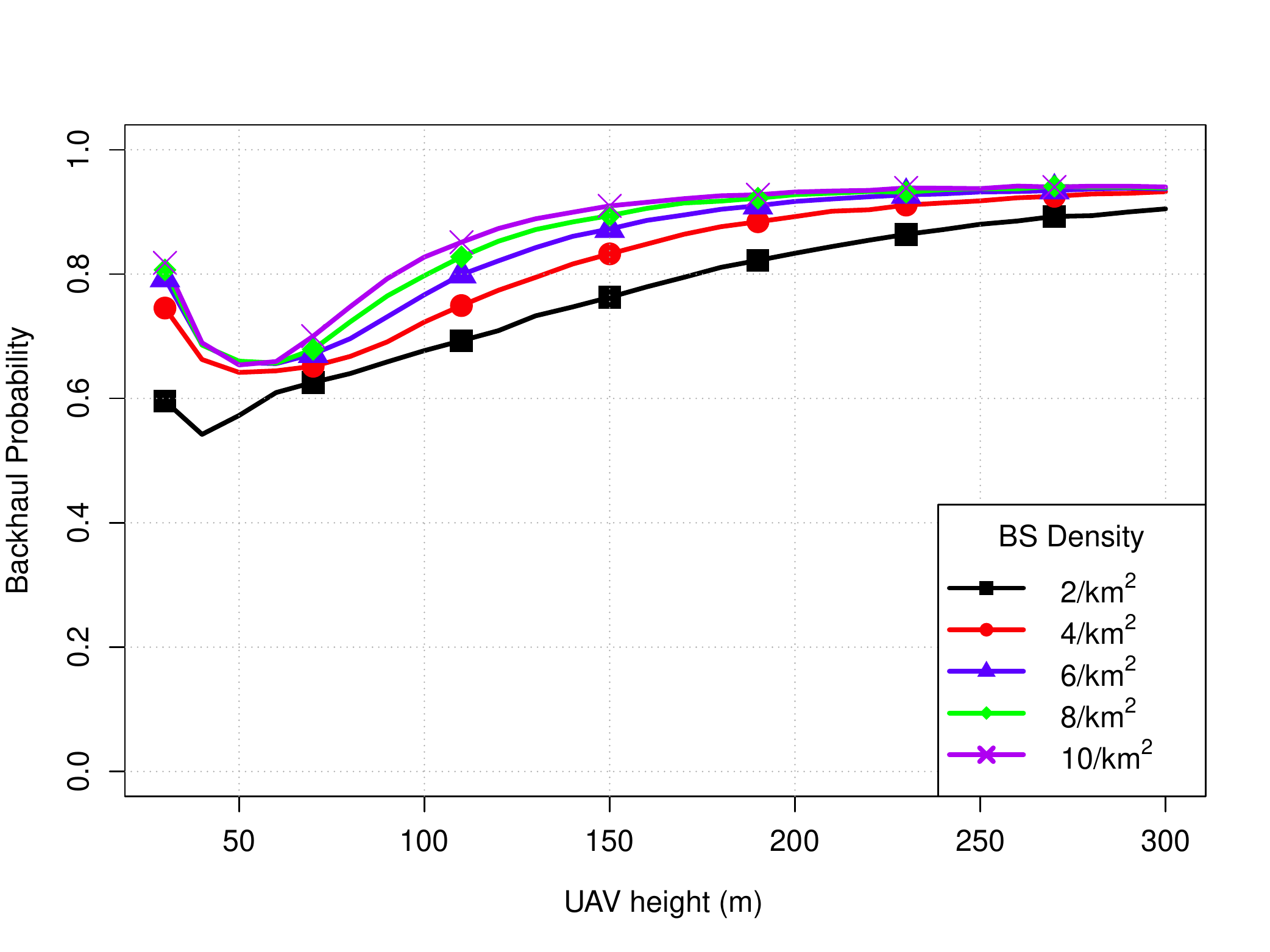}
	\caption{
	Probability of a UAV establishing a backhaul as a function of UAV height and BS density.
	}
	\label{fig:BackhaulDens}

\end{figure}

In \Fig{BackhaulDens} we see the effect the BS density has on the ability of the UAV to establish a backhaul. Higher densities of the BSs correspond to a greater backhaul probability for a given UAV height, as the distance between a UAV and its backhaul BS will on average be smaller, resulting in fewer buildings being in the way and blocking LOS. Additionally, note the effect of the BS beam pattern on the backhaul probability: at very low heights the UAV may find itself inside the main lobe of the BS radiation pattern, but as the UAV height increases it exits this main lobe and instead receives the attenuated sidelobe signals, causing an initial drop in backhaul probability. These results suggest that to maximise the probability of establishing a backhaul the UAV should either operate at very low heights to benefit from BS antenna alignment or should instead increase its height to benefit from high LOS probability to the serving BS.

\begin{figure}[t!]
\centering
	\includegraphics[width=.45\textwidth]{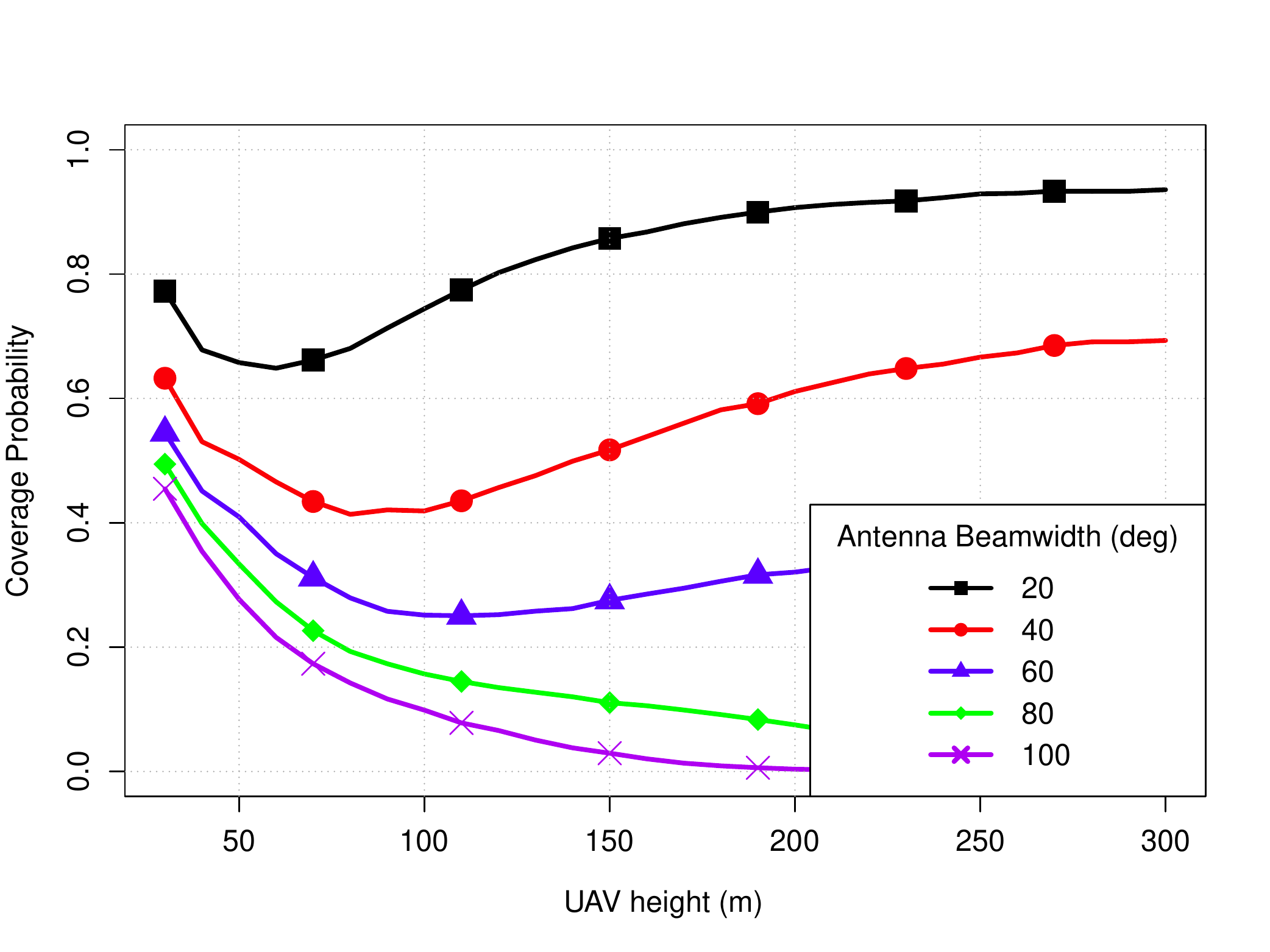}
	\caption{
	Probability of a UAV establishing a backhaul as a function of UAV height and UAV backhaul beamwidth, given a BS density of \unit[5]{$\text{/km}^2$}.
	}
	\label{fig:BackhaulBW}
\end{figure}

In \Fig{BackhaulBW} we demonstrate the effect the UAV backhaul antenna beamwidth has on the backhaul probability. Recall that the beamwidth will affect both the size of the area illuminated by the backhaul antenna as well as the antenna gain. Increasing the beamwidth results in a smaller antenna gain as well as a bigger area, meaning more interfering BS signals can be received by the UAV, which negatively affects the SINR of the backhaul signal. It is clear from the results that the UAV requires a high-quality antenna with a very narrow beam to be able to reliably establish a wireless backhaul into the core network.

\subsection{Coverage probability given a UAV backhaul requirement} 
In this subsection we consider the performance of the UAV network when it has to provide wireless service to a reference user on the ground while ensuring the UAVs can establish a backhaul into the core network. In the previous subsections we have demonstrated how there exists a single UAV height which maximises the coverage probability of the reference user for a given set of UAV parameters; we also demonstrated how a typical UAV can increase the probability of establishing a backhaul by increasing its height. We combine these results to simulate a two-step UAV height optimisation scenario. Initially, we position all UAVs at the height which maximises the coverage probability, as calculated using the mathematical analysis. Each UAV attempts to establish a backhaul through the underlying BS network, and those UAVs that fail to do so increase their height until they either meet the backhaul SINR requirement or until they reach the maximum permissible height. Those UAVs that do not meet the backhaul SINR requirements are considered to be in an outage state and do not provide service to users on the ground. Of the remaining UAVs we determine which are positioned close enough to the reference user to cast their coverage cones over the user position. If there is at least one UAV within range of the reference user, the user associates with it and we calculate whether the reference user link SINR exceeds the threshold $\theta$. The coverage probability for the reference user in this scenario is the joint probability of three events occuring: first, that there is at least one UAV within range of the reference user, second, that at least one of the in-range UAVs meets its backhaul SINR threshold $\theta_B$, third, that the signal received by the reference user exceeds its SINR threshold $\theta$ .

\begin{figure}[t!]
\centering
	\includegraphics[width=.45\textwidth]{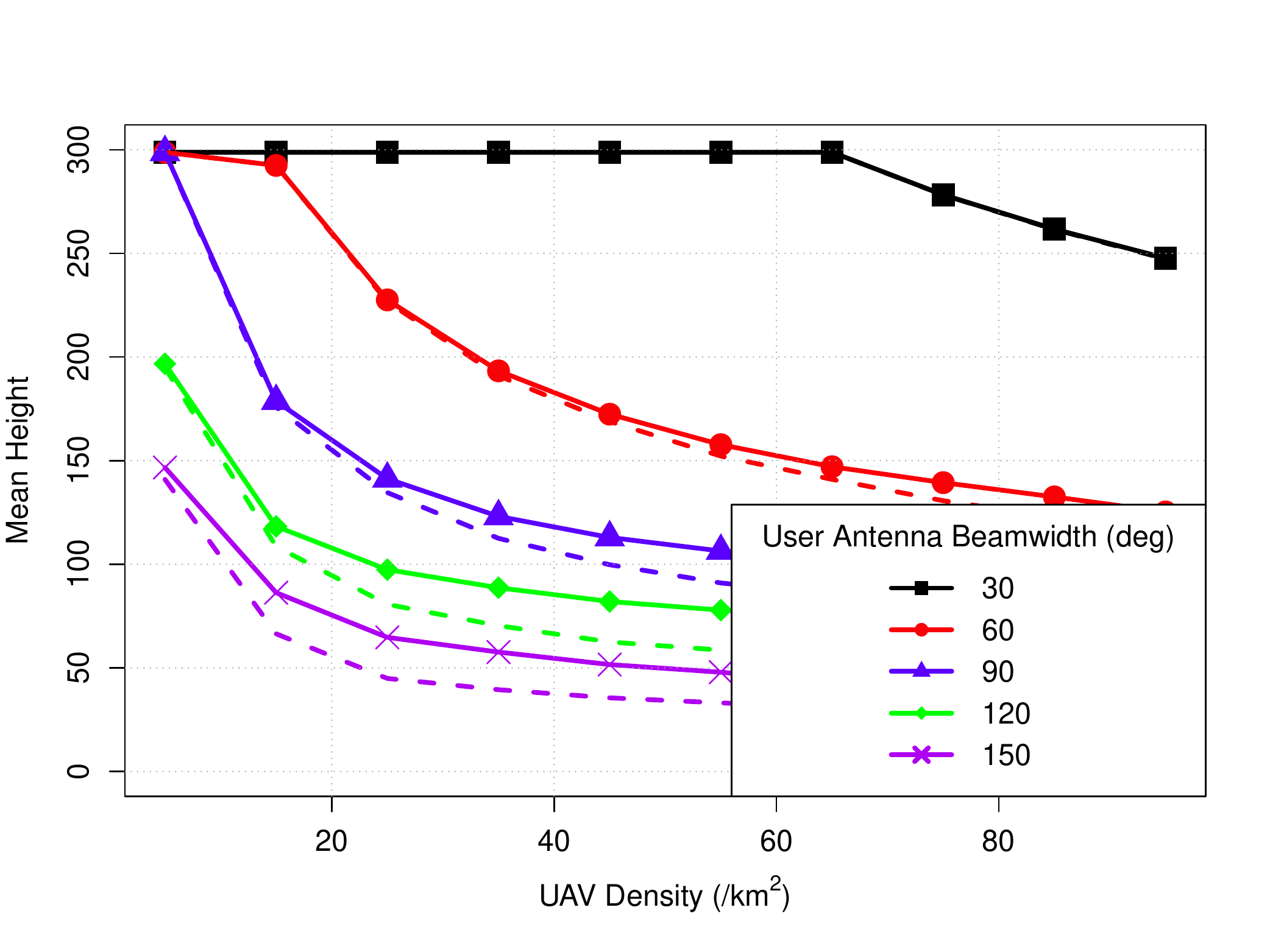}
	\caption{
    Mean height of the UAVs after height optimisation (solid lines with markers) with respect to the backhaul, compared to the optimum height of the UAVs under guaranteed backhaul (dashed lines). UAV backhaul antenna beamwidth is \unit[20]{degrees}.
	}
	\label{fig:MeanHeight}
\end{figure}

In \Fig{MeanHeight} the solid lines with markers indicate the mean height of the UAVs after they carry out their individual height optimisation. Note that we restrict the range of UAV heights to below \unit[300]{m} to stay within the scope of the low-altitude UAV network scenario. We can see that when UAVs are equipped with narrow beamwidth antennas, their optimum height is above \unit[300]{m} for most UAV densities. UAVs are intially positioned at the corresponding optimum height, denoted with the dashed lines. We observe that the mean height of the UAV network after individual optimisation is only \unit[10-20]{m} above this optimum height, since we ensure each UAV does not deviate from the optimum height any more than is absolutely necessary to meet the backhaul requirement. 

\begin{figure}[t!]
\centering
	\includegraphics[width=.45\textwidth]{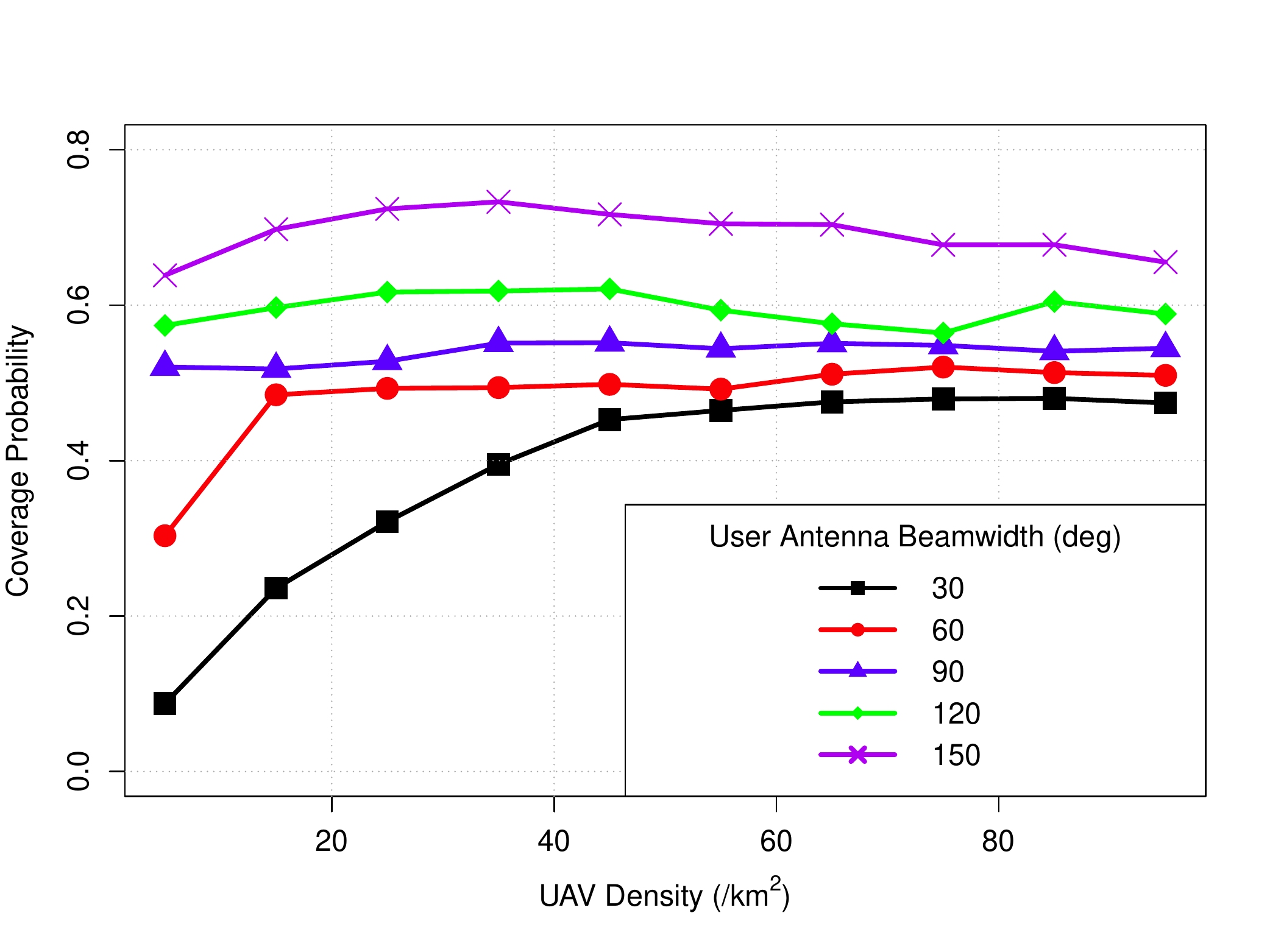}
	\caption{
   Coverage probability for the reference user: joint probability of there existing at least one UAV within range of the user, of that UAV being able to establish a backhaul link with SINR of at least $\theta_B$, and a link to the user with SINR of at least $\theta$.
	}
	\label{fig:JointProb}
\end{figure}

In \Fig{JointProb} we consider the resulting coverage probability of the reference user, after the individual UAVs adjust their heights to ensure they meet their backhaul requirement. We can see that the achievable coverage probability values are comparable to the range of values we reported in the previous subsection when considering all UAVs operating at the same height with guaranteed backhauls.

\section{Conclusion}
In this paper we have used stochastic geometry to model a UAV network in an urban environment, considering UAV network parameters such as density and height above ground, as well as environment parameters such as the building density and building heights. We derived an expression for the coverage probability of the UAV network as a function of these parameters and then verified the derivation numerically, while showing the trade-offs in performance that occur under different network conditions. We then considered the wireless UAV backhaul link and demonstrated that UAVs are able to establish a wireless backhaul using existing terrestrial BSs, provided the UAVs are equipped with high-quality antennas and are able to intelligently adjust their heights. We then considered a scenario where UAV heights are adjusted to meet the UAV backhaul requirement while at the same time attempting to maximise the coverage probability of the end user on the ground. We demonstrated that a UAV network can successfully provide wireless service to a reference user on the ground while also ensuring that the UAVs are able to establish a backhaul signal.

\section*{Acknowledgements}
This material is based upon works supported by the Science Foundation
Ireland under Grants No. 10/IN.1/I3007 and 14/US/I3110. 

\section*{Appendix}

In this appendix we present an analytical expression for a higher order derivative of $\Lc_{I_L}((p\eta(\omega))^{-1}s_L)$. As before, the following results apply to $\Lc_{I_N}(\eta(\omega)^{-1}s_L)$ when $m_L$ and $\alpha_L$ are substituted with $m_N$ and $\alpha_N$. Following \Eq{laplaceFinal} $\Lc_{I_L}((p\eta(\omega))^{-1}s_L)$ can be expressed as a composite function
 
\begin{equation}
\Lc_{I_L}(y(s_L)) = \exp\left(y\big(s_L\big)\right).
\label{eq:comp1} 
\end{equation} 



By expressing the Laplace transform as a composite function we can arrive at a generalised expression for the $n$th derivative of the Laplace transform. We use Fa\`{a} di Bruno's formula to define the $n$th derivative of the composite function \Eq{comp1} with respect to $s_L$ as 

\begin{align}
&\frac{d^{n}\exp\left(y\big(s_L\big)\right)}{ds_L^{n}} = \nonumber \\ 
&\sum\frac{n_k!}{j_1!j_2!...j_{n}!}\exp\left(y\big(s_L\big)\right) \prod\limits_{i=1}^{n}\Bigg(\left(\frac{d^{i}y(s_L)}{ds_L^{i}}\right)/i!\Bigg)^{j_{i}}
\end{align}

where the first sum is over all the tuples that satisfy the sum $\sum\limits_{i=1}^{n}ij_i = n$. The $i$th derivative of $y\big(s_L\big)$ is given as 

\begin{align}
 &\frac{d^{i}y(s_L)}{ds_L^{i}} = -\pi\lambda\sum\limits_{j=\floor*{c_L(\text{L})\sqrt{\beta\delta}}}^{\floor*{\Rcone\sqrt{\beta\delta}}} \Pd_{LOS}(l) \sum\limits_{k=1}^{m_L}\binom{m_L}{k}(-1)^{k+1}\nonumber \\
 &\cdot\bigg(\frac{d^{i} f\left(k,(\Upper^2+\gamma^2),s_L\right)}{ds_L^{i}} - \frac{d^{i} f\left(k,(\Lower^2+\gamma^2),s_L\right)}{ds_L^{i}}\bigg) 
\label{eq:comp2}
\end{align} 

where

\begin{equation}
f\left(k,b,s_L\right) =  b \mbox{$_2$F$_1$}\Big(k,\frac{2}{\alpha_L};1+\frac{2}{\alpha_L};  z(b,s_L)\Big)
\end{equation}

and 

\begin{equation}
z(b,s_L) = -\frac{m_L b^{\alpha_L/2}}{s_L} .
\label{eq:comp3}
\end{equation}

The $i$th derivative of $f\left(k,b,s_L\right)$  with respect to $s_L$ can be obtained using a repeat of Fa\`{a} di Bruno's formula, or we can use [0.430.1] in \cite{Ryzhik_2007}:

\begin{align}
&\frac{d^{i} f\left(k,b,s_L\right)}{ds_L^{i}} = \sum\limits_{q=1}^{i} \frac{U_q}{q!} \frac{d^{q} f\left(k,b,s_L\right)}{dz^q}
\end{align}

where 

\begin{equation}
U_q = \sum\limits_{t=0}^{q-1}(-1)^t \binom{q}{t}z(b,s)^{t}\frac{d^i z(b,s)^{q-t}}{ds_L^i}
\end{equation}

\begin{align}
 &\frac{d^{q} f\left(k,b,s_L\right)}{dz^q} = \nonumber \\ 
 &b \frac{k_{(q)}(2/\alpha_L)_{(q)}}{(1+2/\alpha_L)_{(q)}} \bigg(\mbox{$_2$F$_1$}\Big(k+q,\frac{2}{\alpha_L}+q;1+\frac{2}{\alpha_L}+q; z(b,s_L) \Big)\bigg)
\end{align}

and 

\begin{align}
&\frac{d^i z(b,s_L)^{q-t}}{ds_L^i} = \nonumber \\
&\sum\limits_{e=0}^{i}\sum\limits_{n=0}^{e} (-1)^n(-m_L b^{\alpha_L/2})^e (-m_L b^{\alpha_L/2}/s_L)^{(q-t-e)} \nonumber \\ 
&\cdot\frac{s_L^{(-i-e)}(1+q-t-e)_{(e)}(1+n-i-e)_{(i)}}{n!(e-n)!}
\end{align}

where $(.)_{(a)}$ is the Pochhammer notation for the rising factorial.

\ifCLASSOPTIONcaptionsoff
  \newpage
\fi



\bibliographystyle{./bib/IEEEtran}
\bibliography{./bib/IEEEabrv,./bib/IEEEfull}

\vspace{10mm}
\begin{IEEEbiography}[{\includegraphics[width=1in,height=1.25in,clip]{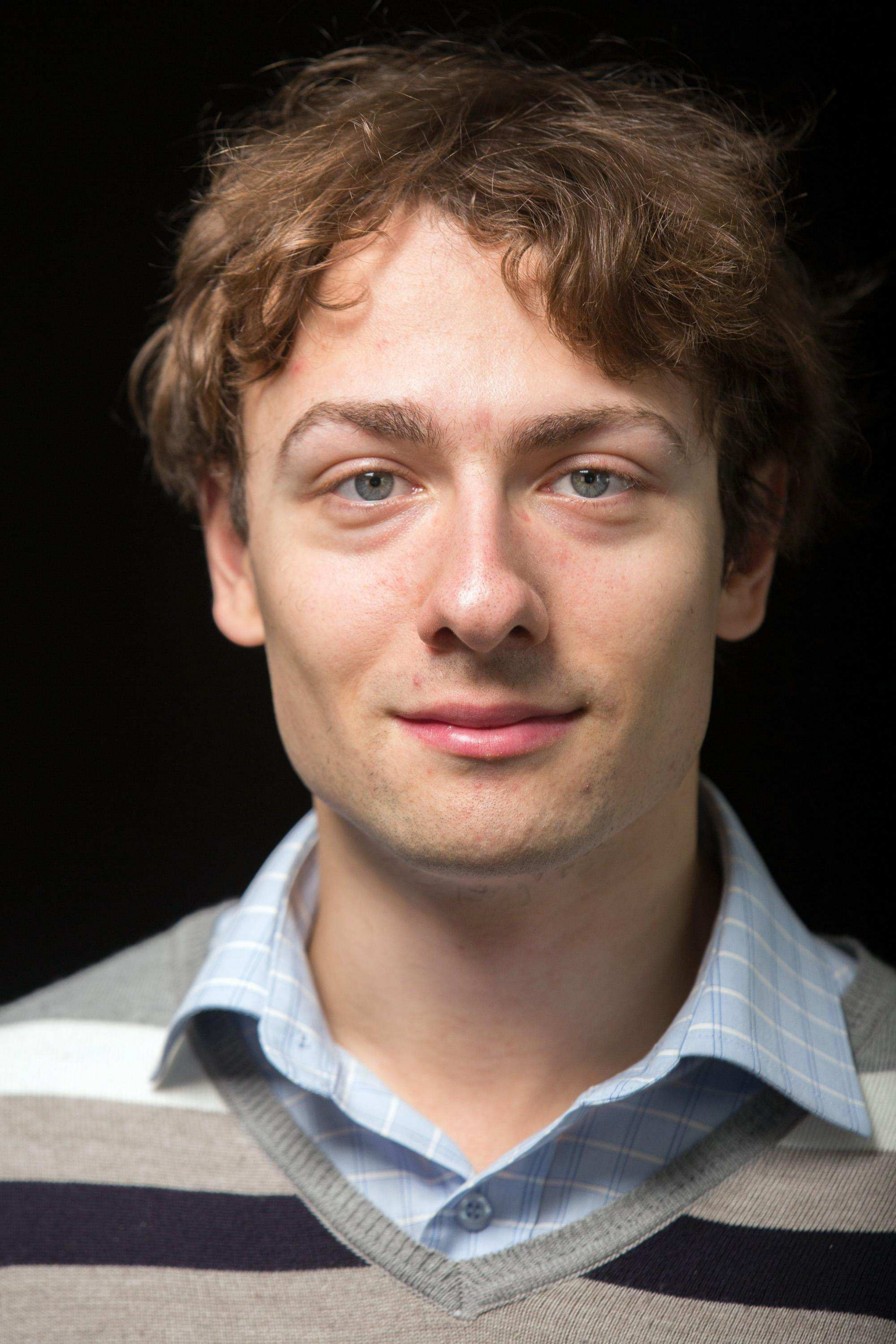}}]{Boris Galkin}
was awarded a BAI and MAI in computer \& electronic engineering by Trinity College Dublin in 2014 and since then has been working toward the Ph.D. degree at CONNECT, Trinity College, The University of Dublin, Ireland. His research interests include small cell networks, unmanned airborne devices and cognitive radios.
\end{IEEEbiography}

\begin{IEEEbiography}[{\includegraphics[width=1in,height=1.25in,clip]{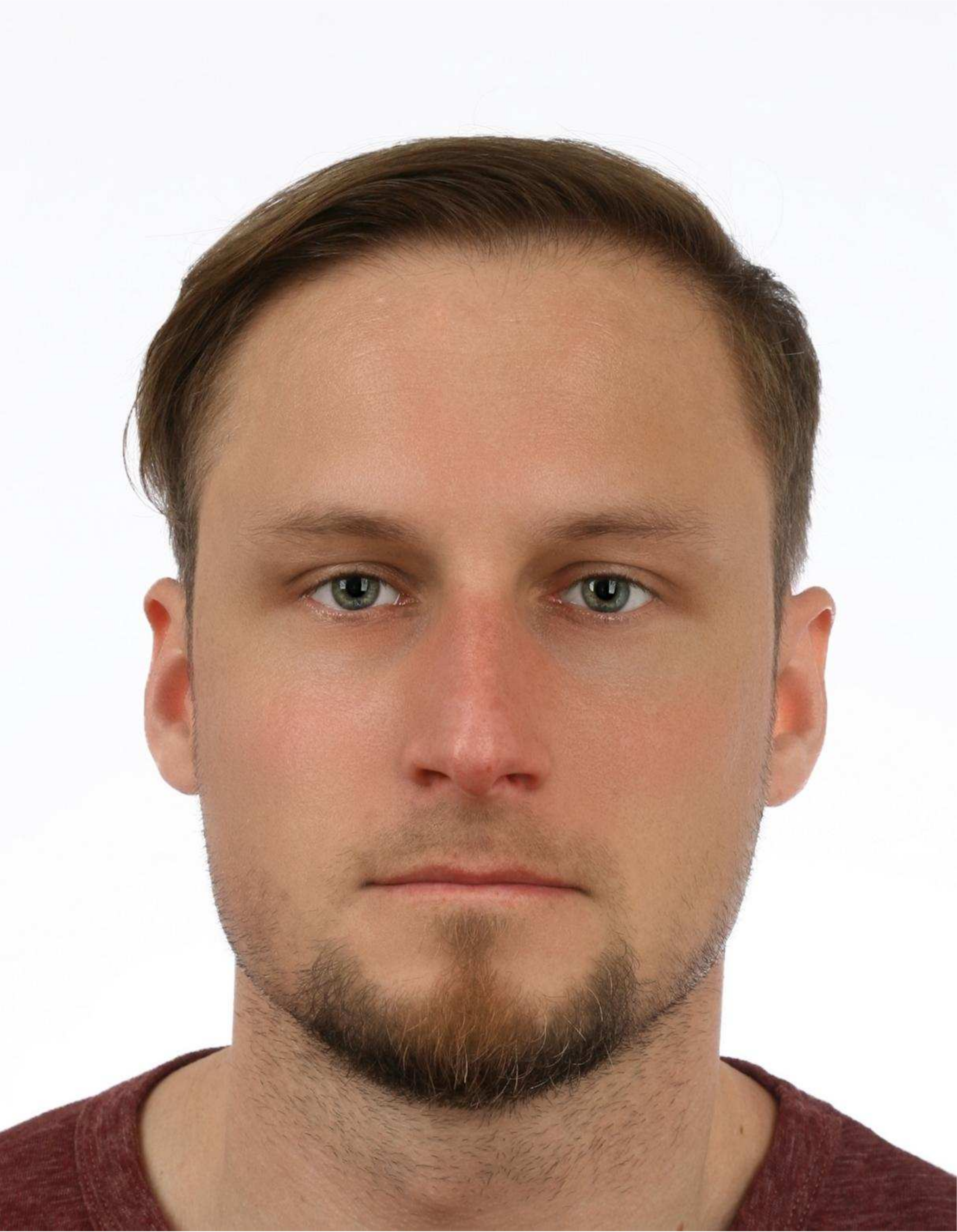}}]{Jacek Kibi\l{}da}
received the M.Sc. degree from Pozna\'n University of Technology, Poland, in 2008, and the Ph.D. degree from Trinity College, The University of Dublin, Ireland, in 2016. Currently he is a research fellow with CONNECT, Trinity College, The University of Dublin, Ireland. His research focuses on architectures and models for future mobile networks.
\end{IEEEbiography}

\begin{IEEEbiography}[{\includegraphics[width=1in,height=1.25in,clip]{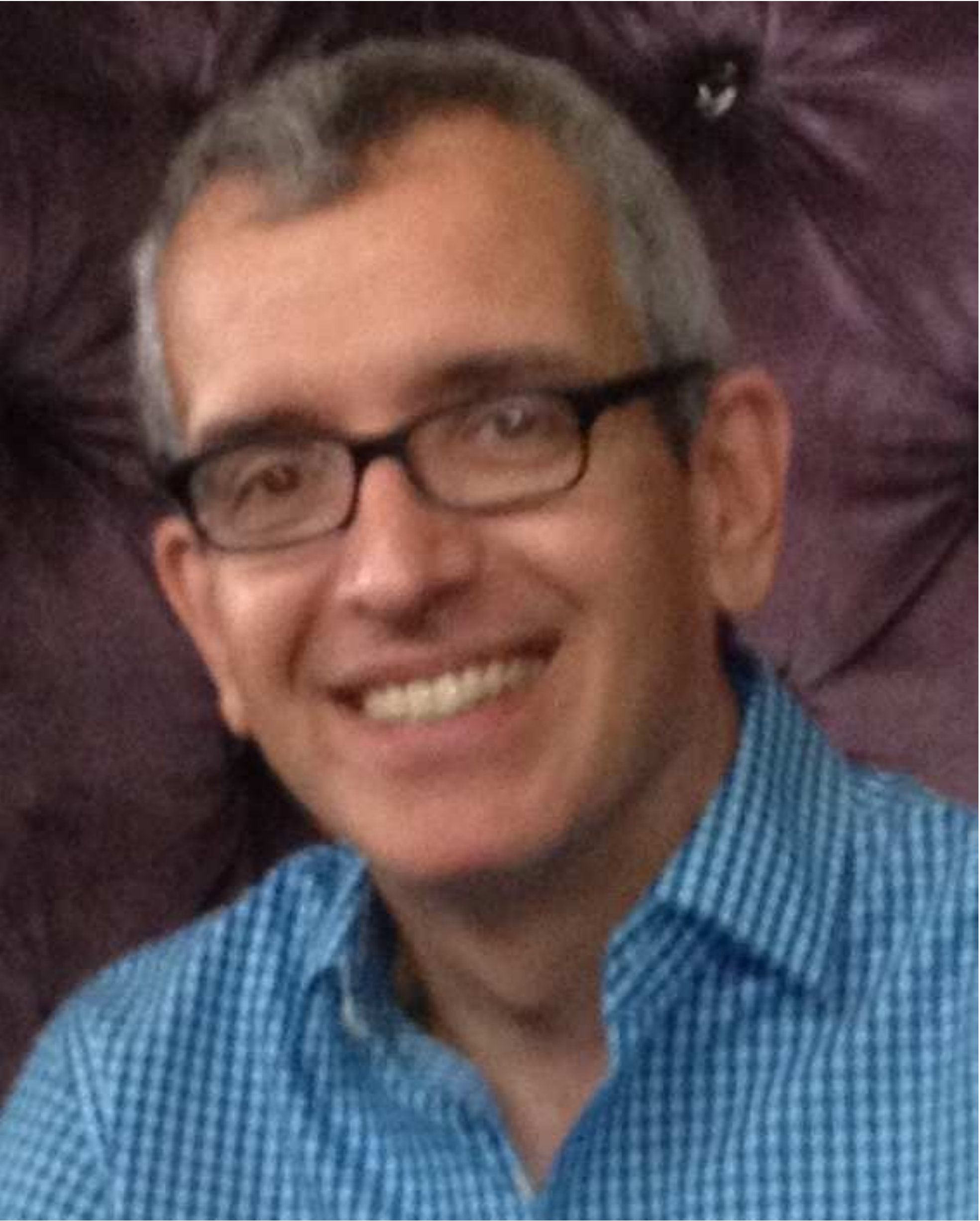}}]{Luiz A. DaSilva}
holds the chair of Telecommunications at Trinity College Dublin, where he is a co-principal investigator of CONNECT, a telecommunications centre funded by the Science Foundation Ireland. Prior to joining Trinity College, Prof. DaSilva was with the Bradley Department of Electrical and Computer Engineering at Virginia Tech for 16 years. Prof DaSilva is a principal investigator on research projects funded by the National Science Foundation, the Science Foundation Ireland, and the European Commission. Prof DaSilva is a Fellow of Trinity College Dublin, an IEEE Communications Society Distinguished Lecturer and a Fellow of the IEEE.
\end{IEEEbiography}
\end{document}